\documentclass[sigconf]{acmart}
\renewcommand\footnotetextcopyrightpermission[1]{} 
\usepackage{caption}
\usepackage{subcaption}
\usepackage{array}
\newcolumntype{P}[1]{>{\centering\arraybackslash}p{#1}}
\newcolumntype{M}[1]{>{\centering\arraybackslash}m{#1}}
\usepackage{setspace}
\usepackage{fancyhdr,graphicx} 
\usepackage[ruled,vlined]{algorithm2e}
\SetAlgoCaptionLayout{raggedright}

\AtBeginDocument{%
  \providecommand\BibTeX{{%
    \normalfont B\kern-0.5em{\scshape i\kern-0.25em b}\kern-0.8em\TeX}}}

\begin{document}

\title{Classification of Encrypted IoT Traffic Despite Padding and Shaping}

\author{Aviv Engelberg}

\affiliation{%
  \institution{School of Computer Science\\Tel Aviv University}
  \city{Ramat Aviv, 69978}
  \country{Israel}
}
  \email{avive1@mail.tau.ac.il}

\author{Avishai Wool}

\affiliation{%
  \institution{School of Electrical Engineering\\Tel Aviv University}
  \city{Ramat Aviv, 69978}
  \country{Israel}
}
  \email{yash@eng.tau.ac.il}

\begin{abstract}\label{Abstract}
It is well known that when IoT traffic is unencrypted it is possible to identify the active devices based on their TCP/IP headers. And when traffic is encrypted, packet-sizes and timings can still be used to do so. To defend against such fingerprinting, traffic  padding and shaping were introduced. In this paper we demonstrate that the packet-sizes distribution can still be used to successfully fingerprint the active IoT devices when shaping and padding are used---as long as the adversary is aware that these mitigations are deployed, and even if the values of the padding and shaping parameters are unknown. The main tool we use in our analysis is the full distribution of packet-sizes---as opposed to commonly used statistics such as mean and variance. We further show how an external adversary who only sees the padded and shaped traffic as aggregated and hidden behind a NAT middlebox can accurately identify the subset of active devices with Recall and Precision of at least 96\%. We also show that the adversary can distinguish time windows containing only bogus cover packets from windows with real device activity, at a granularity of $1sec$ time windows, with 81\% accuracy. Using similar methodology, but now on the defender's side, we are also able to detect anomalous activities in IoT traffic due to the Mirai worm.
\end{abstract}

\maketitle
\pagestyle{plain} 
\section{Introduction}\label{Introduction}
\subsection{Motivation and Background}\label{Motivation and Background}
There is a rapid growth in the number of IoT devices ~\cite{how-many-iot-2}, which leads to security risks regarding inference of user activities from IoT device fingerprinting ~\cite{IoTSense,iot-fingerprinting,IoT-Devices-Recognition}. It is already known that when the traffic is unencrypted it is possible to deduce the active devices in a network with connected smart devices by using features in the TCP/IP headers ~\cite{Beer-Sheva}. Moreover, as IoT vendors move to encrypted traffic, there is metadata that is still available: packet-sizes and timings can also be used in order to infer the identity and activity of IoT devices in a victim network ~\cite{Peek-a-boo,Random-Padding,Ping-Pong,survey}. As a mitigation, prior research has suggested padding the emitted packets with additional bytes and shaping the packet transmission timings to obfuscate the traffic and make it harder for an adversary to identify the devices ~\cite{Random-Padding,STP,shaping-1,Levels-Padding}. Despite such defenses, we show in this paper that even with encryption, padding and shaping applied for the traffic, it is still possible to identify the devices and activities in an attacked network---as long as the adversary is aware that these mitigations are deployed, even if the values of the padding and shaping parameters are unknown. The main tool we use in our analysis is the full distribution of packet-sizes---as opposed to commonly used statistics such as mean and variance.    

Using similar methodology, but now on the defender's side, we are also able to detect anomalous activities in IoT traffic related to possibly compromised smart devices ~\cite{anomaly-ml-1,Mirai-Ref1,anomaly-ml-2}. 

\subsection{Related Work}\label{Related Work}
Prior works ~\cite{Peek-a-boo,Random-Padding,Ping-Pong,survey} have shown that identifying active IoT devices is possible even when all the traffic within the network is encrypted.
Acar et al.~\cite{Peek-a-boo} presented a multi-stage attack for inferring user activities in a smart home. They extracted many binary features related to the metadata and employed different machine learning algorithms for their classification process. 
Trimananda et al.~\cite{Ping-Pong} demonstrated a way to detect devices' ON/OFF events using packet-sizes and directions based signatures. 
St{\"o}ber et al.~\cite{smartphone-fingerprinting} showed that it's possible to identify a smartphone by deriving features like packet-sizes and timings from traffic associated with popular applications like Facebook and WhatsApp. 
Conti et al.~\cite{Analyzing-Android-Encrypted} showed that the actions a user takes on social network and email service apps can be deduced from encrypted traffic on Android mobile devices.

Other papers~\cite{IoTSense,IoT-Devices-Recognition} showed that device type identification can be done by learning different features associated with behavioral patterns of the smart devices.
Some authors~\cite{deep-2,deep-1} utilized deep learning techniques such as Convolutional Neural Networks (CNN) and Long Short-Term Memory (LSTM) to identify devices and recognize specific user commands.
Junges et al.~\cite{Passive-Inference}  developed a signature for every activity and used it to identify the user activities in the network.

It has been suggested by several papers ~\cite{padding-1,padding-2,padding-3} that padding and traffic shaping can help hide the user activities in the network, as they change both the packet-sizes and their rates within the traffic. Such modifications to the packet-sizes and timings prevent most of the prior inference attacks that are based directly on these features. 
Alshehri et al.~\cite{Random-Padding} suggested a method of random padding and showed its efficiency in protecting against packet-size based inference attacks.
Pinheiro et al.~\cite{Levels-Padding} proposed another way to pad packets in the network and showed that it can reduce identification attacks based on packet-size signatures. 
Shaping the traffic to a predetermined rate shared by all the devices can also assist in hiding user activities in the network.
Dziubinsky et al.~\cite{shaping-1} presented how different shaping techniques can significantly enhance the privacy of the network and decrease the adversary confidence in identifying user activities.
In a sequence of papers Apthrope et al.~\cite{apthorpe-1,STP,apthorpe-3,apthorpe-2} also demonstrated that traffic rates can be used to infer various device activities and proposed padding and shaping the traffic as a measure of defense against such attacks. 
However, none of these works examined the case where the \textit{adversary is aware} that padding and shaping are being applied. As we shall see, even if the adversary is just aware that some shaping and padding is occurring, and even without knowing the parameters---the adversary can identify the devices and activities.

On the anomaly detection side, it is already known that detecting suspicious behavior within the IoT traffic is possible even when the traffic is encrypted. Zhang et al.~\cite{anomaly-dfa} have developed a Deterministic Finite Automaton (DFA) model for every smart device activity and used it to identify misbehavior in the traffic, similarly to the Goldenberg-Wool model \cite{gw13} used on industrial control systems' traffic. 
Other papers applied machine learning algorithms in order to identify anomalies in the traffic ~\cite{anomaly-ml-1,anomaly-ml-3,anomaly-ml-2} using statistics such as mean and standard deviation of inter-arrival times as feature vectors. Sridharan et al.~\cite{anomaly-ml-2} implemented modules for both cases of unsupervised and supervised learning and demonstrated how they can identify different types of attacks.
Several works ~\cite{anomaly-deep-2,anomaly-deep-1} have shown how deep learning might also help in discovering when malicious packets are injected into the traffic with  techniques like Convolutional Neural Networks (CNN) and Auto Encoders (AE).

\subsection{Contributions}\label{Contributions}
We show in this paper that the packet-sizes distribution can be successfully used to identify the active devices from encrypted network traffic even when padding and shaping are applied, as long as the adversary is aware that these mitigations are being applied.
We are able to do this either from an internal vantage point, where we can distinguish the device traffic based on Layer 2 data, or from an external vantage point, where we observe traffic that is hidden behind a middlebox (NAT).

While prior works have focused mostly on deriving metadata signatures ~\cite{Random-Padding,Ping-Pong} and statistical metadata information ~\cite{Peek-a-boo} for modeling and identifying devices such as mean and standard deviation, we rely on the full distributions and use measures such as the Cosine distance ~\cite{cosine/kl,cosine-1} and the Jensen-Shannon distance ~\cite{kl-1,cosine/kl} to compare them. 

Specifically, we examined the packet-sizes and inter-arrival times distributions both separately and as a joint distribution, as measures to identify devices in the network.
We show that the packet-sizes and inter-arrival times distributions are not independent, so using both simultaneously does not significantly enhance the adversary's classification power.
We found that the packet-sizes distributions alone is sufficient to identify the active devices within the network.

To evaluate our tests, we examined traffic from a broad number of state-of-the-art IoT devices, taken from available open sources ~\cite{Australian,Ping-Pong}. We obtained excellent results based on a learning time of 3 hours and with only 30 minutes of real-time recorded traffic from the attacked network. 

We demonstrate how a local adversary that has only access to the Layer 2 data exchanged in the LAN can identify the active devices. We further show how an external adversary who only sees the padded and shaped traffic as aggregated by the middlebox can perfectly identify the dominating active devices, and can also infer the subset of active devices with exact identification rate of 75\% and with Recall and Precision of at least 96\%. 

Next we show that the adversary does not need to know the exact padding and shaping parameters---fairly rough estimates suffice; and we present simple methods the adversary can use to estimate the parameters. We also show that the adversary can distinguish time windows containing only cover (i.e., bogus shaped) packets from windows with real device activity, at a granularity of $1sec$ time windows, with 81\% accuracy.

Moreover, in the Appendix we present how the packet-sizes distribution can also be used in the defensive side for achieving novel anomaly detection methods against an active adversary that injects malicious packets into the traffic.
To check our proposed anomaly detection methods, we inject real malicious Mirai attacks packets taken from ~\cite{Attack-Vectors} into the IoT traces. Unlike prior works, that relied upon distribution statistics, using the full distribution as a feature vector our methods do not suffer from false alarms, and have a Precision of 100\% and a Recall of at least 89\% for all the devices.

\section{Preliminaries}\label{Preliminaries}
\subsection{Padding And Shaping}\label{Padding And Shaping}
Obfuscating LAN network traffic relies on two methods: shaping the traffic, i.e., changing transmission times of real packets and injecting cover packets at other times; and padding the emitted packets, meaning the original packet has additional bytes appended.

It was already discussed by Apthorpe et al.~\cite{STP} that one can inject cover packets at a constant rate during the whole active time of the devices. Independent Link Padding (ILP) does exactly that. 
Specifically, it makes sure that all the traffic coming out of the devices is emitted in the same predetermined rate shared by all the devices (injecting cover packets if needed) and that all the packets are either padded to some constant large enough packet-size or their new padded size is drawn from some packet-sizes distribution. When ILP is used over all the devices it becomes impossible to distinguish between them. Hence, this kind of protection results in high privacy as no actual inference can take place. The only possible inference an external adversary might be able to make is the number of active devices.

However, besides providing high privacy against inference attacks, ILP produces a very high bandwidth overhead. Thus, it was suggested to use mitigations that balance between good privacy and reasonable overhead. For that purpose, we consider the Stochastic Padding (STP) method, first suggested by Apthorpe et al.~\cite{STP} 
(see the pseudo code in Algorithm ~\ref{STP}). 
In this kind of padding and shaping, the original traffic is shaped and additional packets are injected in order to create periods of same duration time $T$ with constant predetermined packet rate $R$. Some periods consist of real traffic of the device (mixed with additional possible cover packets) and other periods consist entirely of cover packets. Cover packets are injected stochastically using a predetermined probability $q$. A unique bit flag is included inside the encrypted contents or header of the packet, indicating whether this packet is associated with real traffic or if it is a cover packet and thus should be ignored by the cloud servers and other devices. 

\begin{algorithm}[t]
\SetAlgoLined

injectStart$\leftarrow$ 0

injectEnd $\leftarrow$ 0

 \While{t $\geq$ 0}{
 \If{ t mod T == 0 and randomInjection(q)==True}
        {
        
        injectOffset $\leftarrow$ random(0,T)
        
        \eIf{ t+injectOffset $>$ injectEnd}
        {injectStart$\leftarrow$ t+injectOffset
        
        injectEnd $\leftarrow$ injectStart+T
        }{injectEnd $\leftarrow$ injectEnd+T}
        
        }
 \eIf{ injectStart $<$ t $<$ injectEnd}
        {\For{t' in range(injectStart,injectEnd) } {
        \eIf{ userActivityOccurs(t')}{
        packetSize $\leftarrow$ realPacketSizeAtTime(t')
        }{ 
        packetSize $\leftarrow$  getRandom(distribution)
        }
        paddedSize $\leftarrow$ packetSize+random(1,W)
        
        emitSize(roundUp(t',1/R),paddedSize)
        }
        }{
        \If{ userActivityOccurs(t)}
        {
        injectStart $\leftarrow$ t
        
        injectEnd $\leftarrow$ t+T
        
        \For{t' in range(injectStart,injectEnd) } {
        \eIf{ userActivityOccurs(t')}{
        packetSize $\leftarrow$ realPacketSizeAtTime(t')
        }{ 
        packetSize $\leftarrow$ getRandom(distribution)
        }
        paddedSize $\leftarrow$ packetSize+random(1,W)
        
        emitSize(roundUp(t',1/R),paddedSize)
        }
        }
        }
 }
 \caption{STP($q$, $T$, $R$, $W$, distribution)}
\label{STP}
\end{algorithm}

Apthorpe et al.~\cite{STP} did not specify how one should pad the original packets and how to choose the cover packet-sizes. Several padding methods were already reviewed in previous works such as ~\cite{Random-Padding,Levels-Padding}. In our work we mainly follow the Random Padding method suggested by Alshehri et al.~\cite{Random-Padding}, which pads every packet by a random number of bytes between 1 and a parameter $W$. Random Padding is claimed to provide approximate differential privacy and is useful against attacks that use packet-sizes features ~\cite{Random-Padding,pkt-sizes-identification,Ping-Pong}. Unless indicated otherwise, we follow the default parameter value suggested by Alshehri et al.~\cite{Random-Padding} of $W=80$ and a shaping probability of $q=0.1$ which was found to be an effective balance between privacy and bandwidth overhead against the adversary by Apthorpe et al. ~\cite{STP}. We also investigate the so called ``Level-100'' of padding suggested in ~\cite{Levels-Padding}. 

Besides choosing the padding and shaping, in our version of STP we elected to draw the size (before padding) of the cover traffic packets from the device's real activities packet-sizes distribution.  This makes it more difficult for the adversary 
to distinguish between the cover packets and real packets. 

To prevent STP from introducing latency we make sure that the shaped traffic rate $R$ is higher than the rate of any IoT device's real traffic, and that the time period $T$ is longer than any device's duration of real activity. 
For our tested devices we used $T=1sec, q=0.1$ 
and set the rate $R$ to be $100$ packets per second (pps).

\subsection{Adversary Model}\label{Adversary Model}
We assume that the adversary knows which devices brands may possibly exist in the attacked network: i.e., we assume a  ``closed-world'' model. This is a very realistic assumption as the number of common device manufacturers is not high, and the adversary may be able to discover the installed devices by physical observation or social engineering. 

We also assume that the adversary is aware of the STP algorithm implementation. Combining this assumption with the ``closed-world'' model we get that the adversary is able to simulate the IoT traffic in its own lab, with or without shaping and padding (after estimating the shaping and padding parameters), and use resulting traffic to learn various possible traffic models.

We consider 3 kinds of adversaries: the local adversary, the external adversary ~\cite{Peek-a-boo,STP} and the active adversary ~\cite{anomaly-ml-1,anomaly-ml-3,anomaly-ml-2}.

The local adversary has a passive sniffer on the LAN. Examples for such an adversary are a compromised IoT device or a WiFi eavesdropper that can observe the network traffic devices in the LAN exchange between each other. Assuming that the packets in the attacked network are encrypted by both the WiFi encryption (WPA) and the network protocols (SSL/TLS), the only reliable information that the local adversary has access to are packet-sizes, timings and layer 2 data. Layer 2 data can be used to separate the traffic of IoT devices from each other - however it can be modified easily so we do not rely on it for device identification.

In contrast to the local adversary, an external adversary may be a malicious ISP: the adversary is not within the range of the LAN and cannot see the internal MAC addresses of devices, but only the MAC address of the middlebox. We assume that the middlebox is also performing NAT, so the external adversary sees the aggregated traffic of all the active devices rather than individual per-device traffic flows. Again, assuming that the traffic is encrypted, including layer 3\&4 headers, the adversary only has access to the packet-sizes and timing of the aggregated traffic leaving the middlebox gateway.

The goal of both local and external adversaries is to identify the devices within the network and deduce when they are active, in order to infer user behaviors. Through this identification, they can infer user activity, such as ``if the motion detector is active - then the user is at home and awake''.

Whereas the local and external adversaries are passive and try to infer device identity and activity by sniffing traffic, the active adversary injects malicious packets into the traffic. An example is an IoT device that is taken over by malware such as Mirai ~\cite{Mirai-Ref1,mirai-3,mirai-2}.
Similarly to the local adversary, the active adversary also sees the traffic exchanged by the devices in the LAN. 

\subsection{Methodology}\label{Methodology}
We assume that the adversary analyzes both the normal traffic and the shaped and padded traffic of all the possible IoT devices, under lab conditions. The adversary then extracts features on which to base the device identification in a real-time scenario at a given time in the victim network. 

Since the local and external adversaries' goal is to classify the observed traffic among the possible device brands, similarly to ~\cite{iot-finder}, in our analysis we use confusion matrices to evaluate the performance. Each cell ($i$,$j$) in the confusion matrix is associated with the adversary's learnt model of device $i$ compared to the actual observed traffic associated with device $j$. 
The smaller the value in the cell, the more similar the two models are. When shown graphically we shade the cells in the matrix so darker colors indicate larger values. Therefore, the ideal confusion matrix has the minimal value per column (with the lightest color) on the diagonal---which means that the device is most likely when comparing to its own learnt model than when comparing to all the other devices' learnt models. We define a measure called \textbf{diagonal-rate} of a confusion matrix which measures the fraction of columns that have the minimum value on the diagonal cell. We would like to get a high diagonal-rate, as close to $100\%$ as possible.

Unlike most previous work, we do not rely on statistics like mean and variance in our model. Instead we use the whole distribution of packet-sizes as a feature vector $v$. In other words, for each packet-size $s$ the feature $v[s]$ is the frequency of packets with size $s$.

To compare a learnt model to a tested one we use the Cosine distance ~\cite{cosine/kl,cosine-1}, which is a measure bounded between 0 and 1 based on the cosine of the angle $\theta$ between two vectors associated with the compared distributions' feature vectors:
$$
cosine\_distance(u,v)=1-cos(\theta)=1-\frac{u\cdot v}{\| u\| \| v\| }
$$

While the adversary can record lab traffic for as long as necessary to train the models, it is advantageous to limit the duration of time that the adversary would need to record from the victim network. During preliminary investigation we checked our analysis with different durations of real-time test recording, to achieve the desired classification results. We found that  typically 30 minutes of recorded victim traffic suffice (details omitted). Therefore in all the following sections, unless otherwise specified, we use test recordings of 30 minutes.

\subsection{The Data Corpora}\label{The Data Corpora}
For our testing we consider 14 different smart devices as displayed in Table~\ref{IoT-devices-Table}. The devices' traffic traces are from the open sources of Sivanathan et al.~\cite{Australian} and those of Trimananda et al.~\cite{Ping-Pong}. Every device has recorded network traces lasting many hours. We use separate parts of these traces for learning traffic models as the adversary and for testing against models of other devices.

In addition, we utilize the same traces to develop and evaluate anomaly detection models. To model injected malicious traffic, we use the data from ~\cite{Attack-Vectors} which provides captures from several state-of-the-art malicious attacks traffic, including the Mirai botnet attacks. Mirai malware, as presented in ~\cite{mirai-3,Mirai-Ref1,mirai-2} detected and then compromised IoT devices over the internet by exploiting the fact that many users do not change the default device's vendor credentials. 
In our work, we modeled 3 types of activities that Mirai exhibited: (1) the first stage of infecting a device and connecting to the C\&C malicious server, (2) a SYN flood attack and (3) a DNS attack.

\begin{table}[t]
\centering
\caption{The IoT devices we classified and their data sources}
\begin{tabular}{|M{1.5cm}|M{2.5cm}|M{3cm}|}
\hline

Device Number & Device Name & Origin Dataset \\\hline
1& TP-Link Smart Plug & Sivanathan et al.~\cite{Australian}  \\\hline
2& Belkin Wemo Motion Sensor & Sivanathan et al.~\cite{Australian}  \\\hline
3& Amazon Echo & Sivanathan et al.~\cite{Australian}   \\\hline
4& Netatmo Weather Station & Sivanathan et al.~\cite{Australian}   \\\hline
5& Samsung SmartCam & Sivanathan et al.~\cite{Australian}   \\\hline
6& HP Printer & Sivanathan et al.~\cite{Australian}   \\\hline
7& TP-Link Day Night Cloud Camera & Sivanathan et al.~\cite{Australian}   \\\hline
8& Amazon Plug & Trimananda et al.~\cite{Ping-Pong}  \\\hline
9& D-Link Plug & Trimananda et al.~\cite{Ping-Pong}   \\\hline
10& Rachio Sprinkler & Trimananda et al.~\cite{Ping-Pong}   \\\hline
11& Ring Alarm & Trimananda et al.~\cite{Ping-Pong}   \\\hline
12& Roomba Robot & Trimananda et al.~\cite{Ping-Pong}   \\\hline
13& TP-Link Light Bulb & Trimananda et al.~\cite{Ping-Pong}   \\\hline
14& Belkin Wemo Insight Plug & Trimananda et al.~\cite{Ping-Pong}  \\\hline

\end{tabular}
\label{IoT-devices-Table}
\end{table}

\section{Features that Survive Encryption}\label{Features that Survive Encryption}
\subsection{The Selected Features}\label{Selected Features}
It is already known one can infer IoT device activities using machine learning techniques by extracting many meta-data features out of the tested recording, as shown in several works ~\cite{Peek-a-boo,meidan-1,Ping-Pong}. However, we are interested in a model that uses only features that allow us to identify active devices despite traffic encryption. 

Following ~\cite{Peek-a-boo}, we examined the packet-size and inter-arrival time features, as these features are accessible even when traffic is encrypted.
However, unlike prior work rather than using distribution statistics such as mean, variance, kurtosis etc., we use the full distributions as feature vectors. Packet-sizes range between 54--1594 bytes across all of the examined IoT devices, so we did not need bins. Instead we used a feature vector of at most 1541 frequencies. For inter-arrival times we used bins of width 1 sec, starting from 1 sec up to 108 sec.

To begin with, we checked the classification power of the packet-sizes and inter-arrival times distributions separately and then also as a joint distribution.
In our analysis we modeled the devices from Table ~\ref{IoT-devices-Table} based on a learnt recording of $3$ hours and tested them on random 30 minute recordings from a different day to evaluate how well we modeled the traffic. For the two-features joint distribution we created 2-dimensional bins where each bin is associated with a unique 1 second bin and a specific packet-size bin.

\subsection{Feature Independence}\label{Evaluating The Features}
To evaluate the features, we first tested whether the packet-sizes and inter-arrival times are statistically independent.
To do so we used Pearson's Chi-Squared test of Independence. We observed the contingency table of both packet-sizes and inter-arrival times frequency distribution. Each cell in the contingency table is associated with a unique bin of packet-size and inter-arrival time. In order to use the Chi-Squared test one has to make sure that at least 80\% of the expected values in the contingency table are at least $5$ and that all of them are at least $1$. To reach these conditions we needed to merge bins. We started by creating wider bins: for the inter-arrival times we used bins of 5 sec and for the packet-sizes we used bins of 50 bytes each. Then we iteratively increase the bins size each time with jumps of 5 sec and 50 bytes until they all met the criteria.

In Table ~\ref{chi-independence} we see for example $5$ of the tested devices, and their derived Chi-squared values for the distributions. The table shows that for all these devices the Chi-square value is greater than the critical values at significance level 95\%. Thus, we conclude that the packet-sizes and inter-arrival times distributions are \emph{not} independent: Each device has both a typical size distribution and a typical timing distribution, which implies that using both features has a limited advantage over the use of each separately.

\subsection{Basic Classification Power}\label{Basic Classification Power}
\begin{figure}[t]
  \centering
  \includegraphics[width=\linewidth]{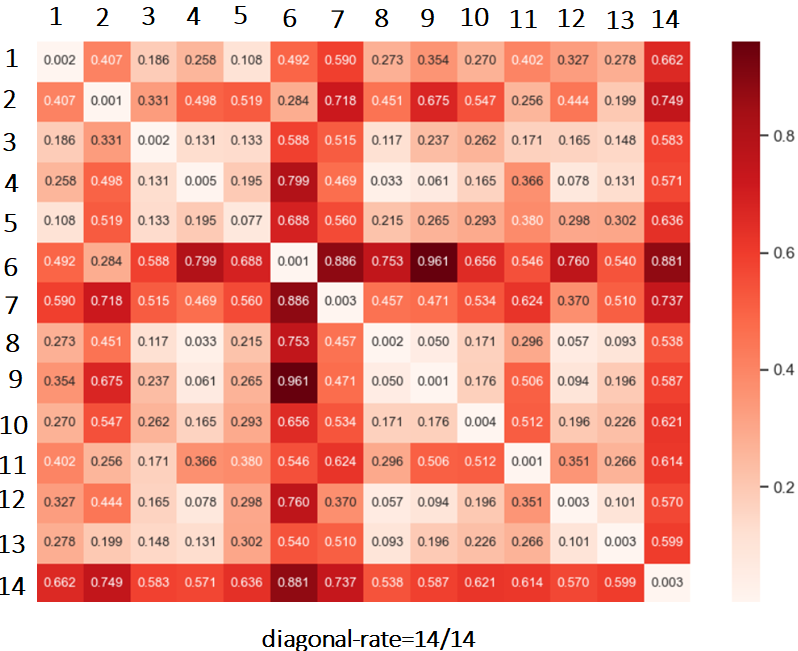}
  \caption{Confusion matrix - Packet-Sizes Model}
  \Description{Confusion matrix that shows the comparison's results of the devices using Packet-Sizes Model, getting a diagonal-rate of 14/14}
  \label{Confusion matrix - Packet-Sizes Model}
\end{figure}

\begin{figure}[t]
  \centering
  \includegraphics[width=\linewidth]{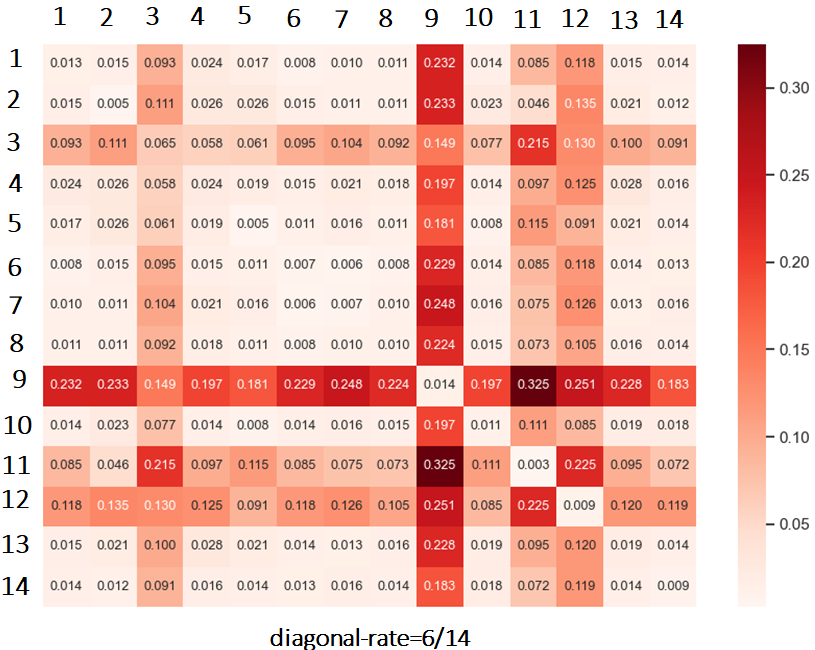}
  \caption{Confusion matrix - Inter-Arrival Times Model}
  \Description{Confusion matrix that shows the comparison's results of the devices using Inter-Arrival Times Model, getting a diagonal-rate of 6/14}
  \label{Confusion matrix - Inter-Arrival Times Model}
\end{figure}

Next we evaluated the classification power of the features in the external adversary model in the scenario of one single dominating device when STP is applied (for $W=80$, $q=0.1$) in the attacked network. We computed the confusion matrix using each distribution separately (Figures ~\ref{Confusion matrix - Packet-Sizes Model},~\ref{Confusion matrix - Inter-Arrival Times Model}) and using the joint distribution (Figure ~\ref{Confusion matrix - (packet-sizes,Inter-Arrival Times) Model}). 
As discussed in Section ~\ref{Methodology}, a cell $(i,j)$ in the confusion matrix represents the cosine distance measure between the learnt distribution of device $i$ and tested distribution of device $j$.
Figure ~\ref{Confusion matrix - Packet-Sizes Model} shows that the packet-sizes distribution is a good feature to differentiate devices with a diagonal-rate of 14/14. Each smart device has quite a unique set of packet-sizes associated with it, derived both from the brand of that device and from the activities it performing.
On the other hand, using only the inter-arrival times distribution as seen in Figure ~\ref{Confusion matrix - Inter-Arrival Times Model} is less effective as the diagonal-rate is only $6/14$ which means that this method would incorrectly classify 8 of our devices. 

When comparing Figures ~\ref{Confusion matrix - Packet-Sizes Model} (packet-sizes distribution) and ~\ref{Confusion matrix - (packet-sizes,Inter-Arrival Times) Model} (joint distribution) we see that both have a perfect diagonal-rate of $14/14$, with a slight advantage for the joint distribution: it has darker  (larger value) cells off the diagonal, which means that the devices there are distinguished more easily. However the difference between the power of the packet-size method and the joint size-and-time distribution method is small, thus to simplify our models in the following sections we focus only on the packet-sizes distributions. 

\begin{figure}[t]
  \centering
  \includegraphics[width=\linewidth]{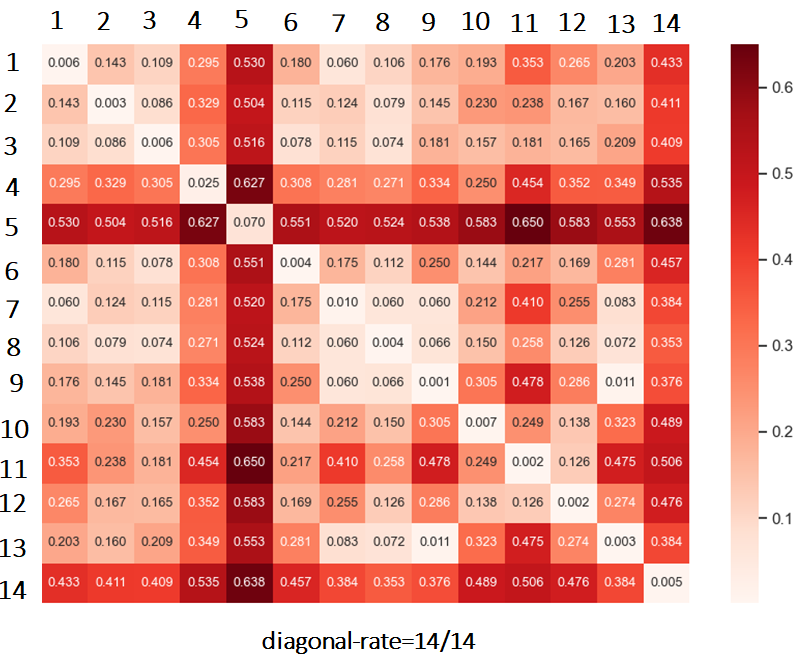}
  \caption{Confusion matrix - Joint distribution Model}
  \Description{Confusion matrix that shows the comparison's results of the devices using Joint distribution Model, getting a diagonal-rate of 14/14}
  \label{Confusion matrix - (packet-sizes,Inter-Arrival Times) Model}
\end{figure}

\begin{table*}[t]
\small
\centering
\caption{Chi-Squared Test of Independence between the packet-sizes and inter-arrival times distributions}
\begin{tabular}{|M{2cm}|M{1.8cm}|M{1.8cm}|M{2.2cm}|M{2.2cm}|M{2.6cm}|M{2cm}|}
\hline

Device Name &Inter-arrival time bins & Packet-size bins &
Percentage of expected values > 5 & Degrees of freedom
&Critical value at significance level 95\% & Chi-squared Value  \\\hline
TP-Link Plug & 100 & 250  & 83 & 6& 12.59 & 16.21\\\hline
Belkin Motion & 50 & 1000  & 87 & 1& 3.84 & 5.4\\\hline
Amazon Echo & 10 & 1000  & 100 & 2& 5.99 & 8.37\\\hline
Netatmo Weather & 100 & 250  & 90 & 4& 9.49 & 11.1\\\hline
Samsung Camera & 5 & 500  & 100 & 15& 25.1 & 26.89\\\hline

\end{tabular}
\label{chi-independence}
\end{table*}

\section{The Local Adversary Model}\label{Local Adversary Model}

As discussed in Section \ref{Preliminaries}, the local adversary can filter the observed padded and shaped traffic by the MAC addresses and attempt to identify which MAC address belongs to which suspected smart device.
To evaluate the adversary's ability to do so, we classify the devices based on the packet-sizes distribution. In this section we assume that the adversary knows the shaping and padding parameters - in Section \ref{Identifying the shaping and padding parameters} we explore whether the adversary needs to know them accurately.

\begin{figure}[t]
  \centering
  \includegraphics[width=\linewidth]{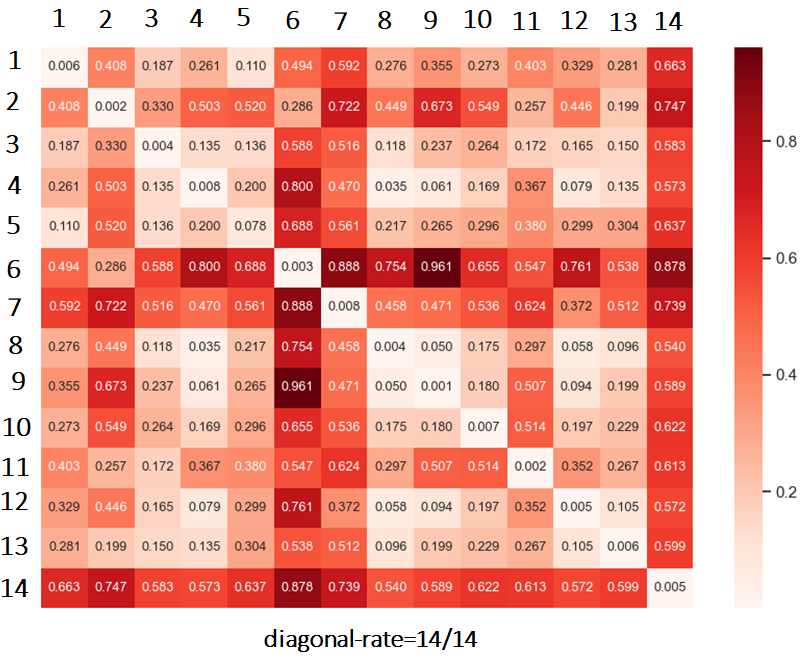}
  \caption{Confusion matrix - Local Adversary}
  \Description{Confusion matrix that shows the comparison's results of the devices in the Local Adversary Model, getting a diagonal-rate of 14/14}
  \label{Confusion matrix - Local Adversary}
\end{figure}

As mentioned in Section ~\ref{Preliminaries} we use the parameters $W=80, T=1sec, R=100pps$ , $q=0.1$ (recall Algorithm ~\ref{STP}).
The adversary trains his model based on 3 hours of recorded traffic that is shaped and padded with STP using those parameters, and then a shaped and padded sample (from another day) is classified. We repeated this experiment with 10 samples and 10 training traces.
The confusion matrix in Figure ~\ref{Confusion matrix - Local Adversary} shows that the adversary can successfully determine which device is associated with which MAC address: the diagonal-rate is $14/14$. Thus, even when shaping and padding take place in the LAN we are able to identify the devices based solely on the packet-sizes distribution.

\section{The External Adversary Model}\label{External Adversary Model}
\subsection{Identifying an Active Dominating Device}\label{Identifying Active Dominating Device}
As discussed in Section ~\ref{Preliminaries} and in ~\cite{Peek-a-boo,STP}, the external adversary has access to the metadata of the aggregated traffic coming out of the home gateway which acts as a NAT middlebox. 
We consider the same shaping and padding parameters as before  $W=80, T=1sec, R=100pps$ , $q=0.1$.
Therefore the external adversary's goal is to identify which device, or group of devices, is active, despite their traffic being aggregated. 
The simplest case is when only one dominating device is active at a specific time. 
In this case, the analysis is identical to that of the local adversary, and the results of Figure~\ref{Confusion matrix - Packet-Sizes Model} (see Section~\ref{Basic Classification Power}) show:
similarly to the local adversary, the external adversary can also tell which dominating device emits packets at a specific time. 

\subsection{Identifying the Number of Active Devices}\label{Identifying Number Of Active Devices}

When multiple devices are simultaneously active in the victim network, a relatively basic goal for the external adversary is to determine their number. We suggest to do this based on the packet rate produced by each combination of active devices. The adversary can, in advance, try all possible combinations of learnt active devices and calculate the expected packet rate as a function of their number. 
When it has all the average packet rates for each possible number of active devices it can set thresholds between them to determine the number of active devices. Then, at attack time, it would record the victim's traffic coming out of the middlebox gateway  and measure the packet rate. By comparing it to the traffic thresholds, it can estimate the number of active devices. 

\begin{figure}[t]
  \centering
  \captionsetup{justification=centering}
  \includegraphics[width=\linewidth]{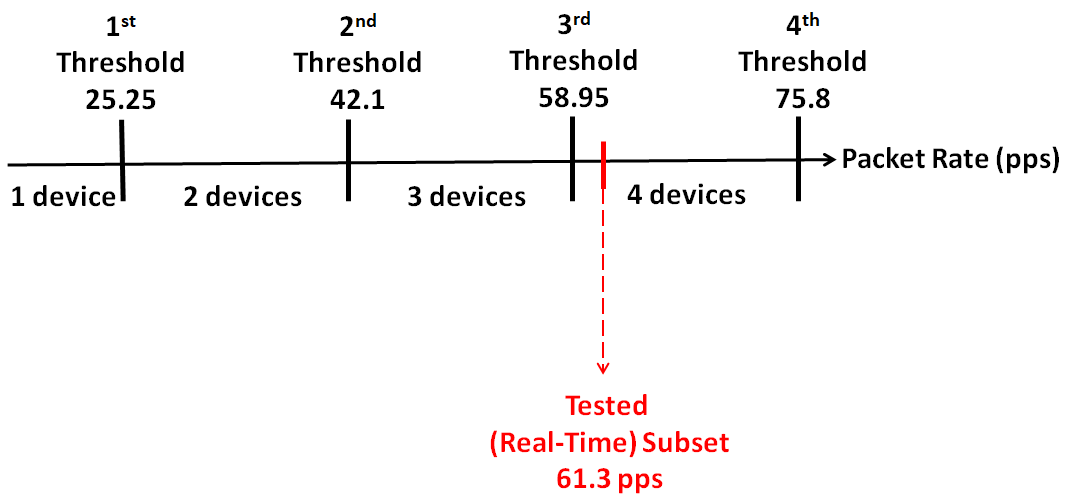}
  \caption{Example of the Classifier finding the number of active devices. If the adversary observes traffic with rate of 61.3 packets per second (pps) it concludes that 4 devices are active}
  \Description{Example that shows how the classifier  estimate the number of active devices using the thresholds}
  \label{Thresholds}
\end{figure}

To develop our model, we take 30 minutes recording of every combination of active devices and compute the average packet rate for each possible number of active devices (between 1 to 14). We set the thresholds to be the mid-points between averages. In Figure ~\ref{Thresholds} we can see an example of the first 4 thresholds in a network of the 14 devices we evaluated.

To evaluate the model, we tested it on a different 30 minutes recording from the attacked network - each time of a different combination of active devices, for all the possible combinations - and compared the number estimated by the model to the actual number of active devices. We repeated this experiment 10 times, each time with a different trace for learning and testing.

Our results are that the estimated number is correct with accuracy of 63\%, and that the classifier always estimates the number of active devices to be the actual number of active devices plus/minus~1. 

\subsection{Detecting Subset of Active Devices}\label{Detecting Subset of Active Devices}

\subsubsection{Methodology}\label{External Adversary-Methodology}
The external adversary's goal is to estimate the subset of active devices as closely as it can to the true subset. Analogous to the standard definitions of Precision and Recall,
in the subset-detection scenario we define ``Precision'' to mean the fraction of 
truly active devices among the estimated set, and we define ``Recall'' to mean the fraction of estimated-to-be-active devices among the truly active devices. Our presented results in Table \ref{short+long checks} refer to the average Precision and Recall over many tests.
We also define the ``Exact Identification'' to mean the fraction of tests in which the estimated subset was precisely the active set (i.e., tests in which Precision $=$ Recall $=100\%$). 

\subsubsection{Full Comparison Check}

The most naive approach is to compare the tested subset's packet-sizes distribution to all possible subsets' packet-sizes distributions. 
The one whose similarity to the tested subset is the highest would be the estimated subset of devices. With $n$ candidate devices this approach would require comparing to $2^n$ distributions, which makes it prohibitive for $n$ larger than a handful of possible devices.

However, in Section ~\ref{Identifying Number Of Active Devices} we saw that the estimated size by the Number Of Active Devices Classifier is always at most off by 1 than the true subset's size. We can use this observation to reduce the time complexity without losing accuracy: it's enough to check only the subsets of devices whose size is at most off by 1 than the estimated size. With $n=14$, using this optimization to only try 3 possible sizes of subsets we reduced the average time of identifying the subset by a factor of 10: from 30sec down to 3sec.

Table ~\ref{short+long checks} (right columns) shows the results for different number of candidate devices in the  network. 
We see that we get Precision and Recall of at least 96\% and Exact Identification of at least 75\% for any number of devices in the network.

\subsubsection{Fast Scores Based Check (FSBC)}

Despite the optimization of only testing subsets of the estimated number $\pm 1$, the Full Comparison Check still takes time that is exponential in $n$ at test time, so it is only effective for small values of $n$. In this section we suggest a faster, non-exponential test-time algorithm we call the Fast Scores Based Check (FSBC) - see Algorithm~\ref{fsbc}. This algorithm also relies on knowing the number of active devices $\pm 1$.

The basic building block of FSBC, given an aggregate packet-sizes distribution and the learnt packet-sizes distribution of a specific device $i$, is to give a score indicating how likely it is that $i$ is active in the aggregate distribution. Initially all the devices start with a default score, arbitrarily set to 1 (minimal score is 0).

FSBC first checks what packet-sizes are unique to device $i$, i.e., packet-sizes that device $i$ emits but other candidate devices do not (this check can be done in advance during the learning phase). Because it might happen that unique sizes with low frequencies of some device in the learnt recording would still occur in a different device's distribution in the tested aggregated recording, for each device FSBC only considers the unique size with the top frequency. If the top-frequency unique size of device $i$ does not occur in the tested aggregated recording, this is a strong indication that $i$ is not active, thus its score is decreased.

Next, the adversary identifies for each device $i$ its common packet-sizes: these are the $f_1$ most frequent packet-sizes, for parameter $0\% \leq f_1 \leq 100\%$. Then, for each common size with frequency $s$, FSBC checks whether there are at least $f_2 \cdot s$ packets with that size in the tested aggregated recording, for another parameter $0\% \leq f_2 \leq 100\%$. If device $i$'s most common packet-sizes occur less frequently in the aggregated traffic this is another indication that $i$ is not active, and its score is decreased.

Algorithm \ref{fsbc} calculates the score as above for all $n$ candidate devices. At the end, it discards the devices with the lowest scores until it remain with only the top-scoring estimated number of devices by the Number Of Active Devices Classifier from Section \ref{Identifying Number Of Active Devices}. 

To evaluate the Fast Scores Based classifier we tested it over all possible subsets of different devices, and repeated this 10 times, each time with a different trace of learning and testing.
To set the values of $f_1, f_2$ we ranged their values from $0\%$ to $100\%$ with jump of $5\%$ and compared the results of the 10-times experiments each time with a different value of $f_1, f_2$. We got the best results for $f_1=80\%, f_2=90\%$.

Table ~\ref{short+long checks} (left columns) shows that the Precision, Recall and Exact Identification are all larger when the number of devices is smaller. We get Precision and Recall of at least 74\% for any number of devices. 

Comparing the results of FSBC to the Full Comparison Check in 
Table ~\ref{short+long checks} we see that with 5 IoT devices, both methods have Precision and Recall of 97--98\%, but FSBC has a low rate of Exact identification for subsets of 10 devices or more.

However, the Full Comparison Check comes at a cost: while the Fast Scores Based Check has an initialization time of about 5 seconds to learn the common and unique packet-sizes for network with 14 devices, the Full Comparison Check requires an initialization time of about 30. Also, the average time of identifying the subset (after the initialization is completed) is also much faster for the Fast Scores Based Check: about 10ms while it's almost 3 seconds for the Full Comparison Check. This high ratio between the initialization and identification times of the Full Comparison and the Fast Scores Based checks remains the same for any number of devices in the network. Thus we conclude that FSBC gives the adversary a very good estimate of the subset of active devices at a fraction of the computational cost of the Full Comparison Check.

\begin{algorithm}[t]
\SetAlgoLined
scoresList $\leftarrow$ [\hspace{0.05cm}]

 \For{\textnormal{d} in \textnormal{distributions} }{
score $\leftarrow$ 1
 
 uniqueSize $\leftarrow$ extractTopUniqueSize(d,distributions)
 
 commonSizesList $\leftarrow$ extractCommonSizes(d,$f_1$)

 total $\leftarrow$ 1 + len(commonSizesList)

 \If{ \textnormal{uniqueSize} not in  \textnormal{testedSubset}}
        {score $\leftarrow$ score - 1/total}
    
    \For{\textnormal{packetSize} in \textnormal{commonSizesList} }{
        \If{ \textnormal{testedSubset[packetSize]}$< f_2\cdot$\textnormal{d[packetSize]}}{score $\leftarrow$ score - 1/total}
    }
    scoresList.append(score)
 }
estimatedSize$\leftarrow$numberOfActiveDevicesClassifier(testedSubset)

estimatedSubset$\leftarrow$takeTopScoresDevices(scoresList,estimatedSize)

\Return estimatedSubset

 \caption{FSBC(distributions[\hspace{0.05cm}], $f_1$, $f_2$, testedSubset)}
\label{fsbc}
\end{algorithm}

\begin{table}[t]
\scriptsize
\centering
\caption{Fast Scores Based + Full Comparison Checks}
\begin{tabular}{|M{0.8cm}|M{0.6cm}|M{0.65cm}|M{1.3cm}|M{0.6cm}|M{0.65cm}|M{1.3cm}|}
\hline

Number
of 
Devices&\multicolumn{3}{c|}{Fast Scores Based Check}&\multicolumn{3}{c|}{Full Comparison Check}\\\cline{2-7}
&Recall
(\%)&Precision
(\%)&Exact 
Identification
(\%)&Recall
(\%)&Precision
(\%)&Exact 
Identification
(\%)\\\hline
5 & 97 & 97 & 74   & 98 & 98 & 81\\\hline
10 & 79 & 83 & 18  & 97 & 97 & 77\\\hline
14 & 74 & 76 & 4   & 97 & 96 & 75\\\hline
\end{tabular}
\label{short+long checks}
\end{table}

\subsection{Results with Level-100 Padding}\label{Results with Levels Padding}

\begin{figure}[t]
  \centering
  \includegraphics[width=\linewidth]{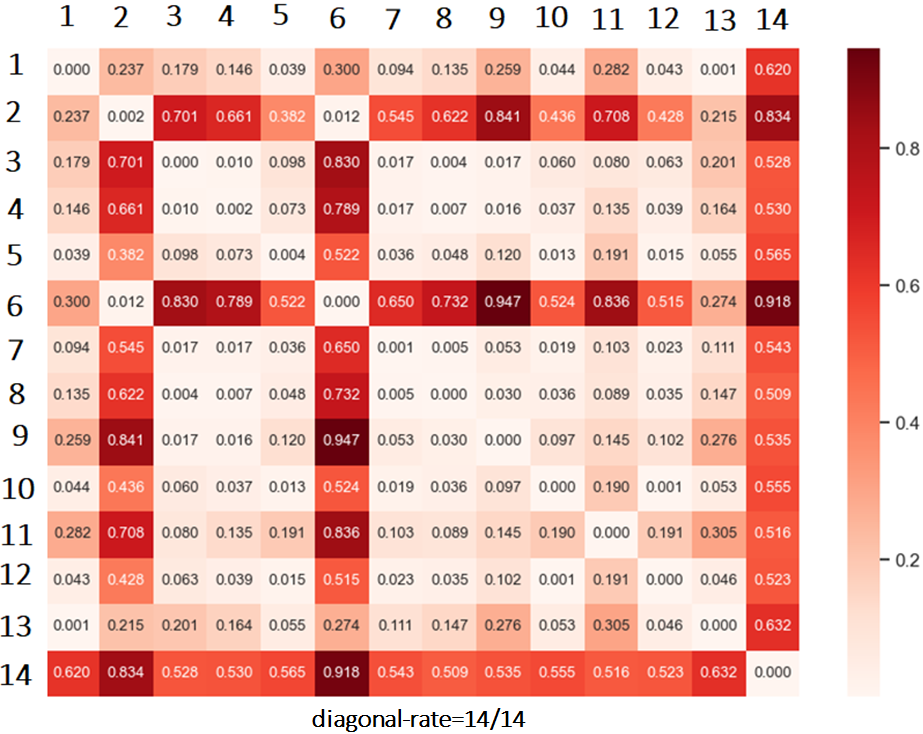}
  \caption{Confusion matrix - Level-100 Padding}
  \Description{Confusion matrix that shows the comparison's results of the devices in the case of Level-100 Padding, getting a diagonal-rate of 14/14}
  \label{Confusion matrix - Level 100 Padding}
\end{figure}

Another padding method, suggested by Pinheiro et al.~\cite{Levels-Padding}, is Level-100 Padding (see Table ~\ref{Level 100 Padding}). The motivation to use this padding is that it's considered to have lower bandwidth overhead. In addition, Level-100 padding is considered to offer better privacy as more packets share the same packet-sizes. We used the same methods as in Sections ~\ref{Identifying Active Dominating Device}--~\ref{Detecting Subset of Active Devices} in order to estimate the efficiency of this measure of defense. As seen in Figure ~\ref{Confusion matrix - Level 100 Padding}, in the case of a single dominating device we still manage to identify it correctly both in the case of local adversary and external adversary. 

We then tested the effect of Level-100 padding on the Number Of Active Devices Classifier (Section \ref{Identifying Number Of Active Devices}) and found out that it didn't change the results:
As expected, Level-100 padding changes the way we pad the packets but not the number of sent packets, hence it does not reduce the ability to infer how many devices are currently active at the victim's network. 

In addition, when trying to detect subsets of active devices as before, we found out that Level-100 padding has an effect mainly on the exact identifying of the active subset, but nonetheless we still managed to obtain quite high measures of Precision and Recall (about 76\% - see Table \ref{Level 100 padding - Fast Scores Based + Full Comparison Checks}), meaning that an adversary can still get high confidence of the active devices in the victim network. The only measure that is significantly impacted by Level-100 padding is the exact identification which dropped to about 10\%: this makes sense as with this kind of padding many packets share the same size and thus detecting the exact subset of active devices becomes considerably more difficult.

\begin{table}[t]
\centering
 \caption{Level-100 Padding}
 \begin{tabular}{c c c c} 
 \hline
 Original Size(s) & Padded Size \\ [0.5ex] 
 \hline
 $s$ $\leq$ 100 & 100  \\ 
 100< $s$ $\leq$ 200 & 200 \\
 200< $s$ $\leq$ 300 & 300  \\
 300< $s$ < 999 & random size from $s$ to 1000  \\
 999$\leq$ $s$ $\leq$ 1399 & random size from $s$ to 1400  \\ 
 1400$\leq$ s $\leq$ 1600 & 1600  \\ [1ex] 
 \end{tabular}
 \label{Level 100 Padding}
\end{table}

\begin{table}[t]
\footnotesize
\centering
\captionsetup{justification=centering}
\caption{Level-100 padding - \\Fast Scores Based + Full Comparison Checks}
\begin{tabular}{|P{0.7cm}|P{0.75cm}|P{1.4cm}|P{0.7cm}|P{0.75cm}|P{1.4cm}|}
\hline

\multicolumn{3}{|c|}{Fast Scores Based Check}&\multicolumn{3}{c|}{Full Comparison Check}\\\cline{1-6}
Recall
(\%)&Precision
(\%)&Exact 
Identification
(\%)
&Recall
(\%)&Precision
(\%)&Exact 
Identification
(\%)\\\hline
54 & 55 & 1   & 77 & 76 & 8\\\hline
\end{tabular}
\label{Level 100 padding - Fast Scores Based + Full Comparison Checks}
\end{table}

\begin{figure*}
     \centering
     \begin{subfigure}[t]{0.45\textwidth}
         \centering
         \captionsetup{justification=centering}
         \includegraphics[width=\textwidth]{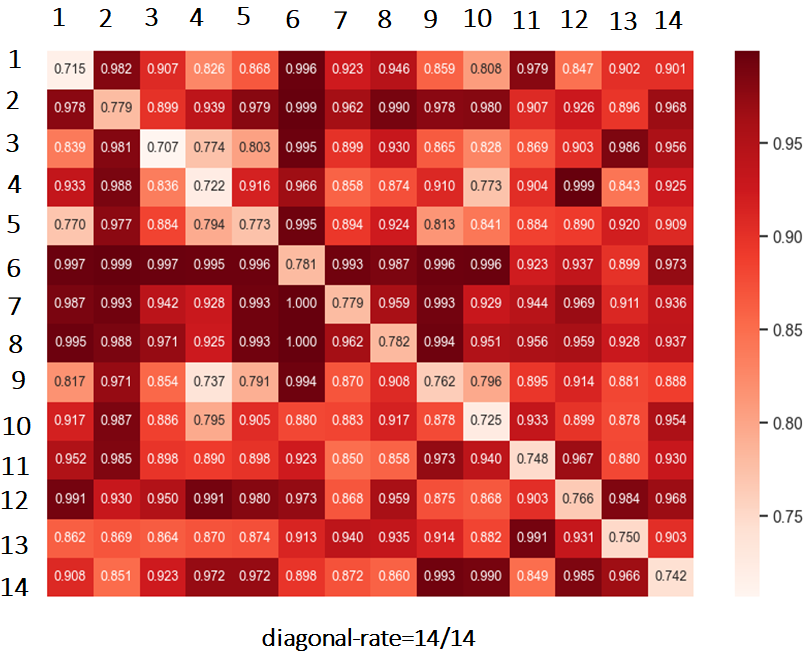}
          \caption{Confusion matrix - Learning with W=60, Testing with W=80}
         \Description{Confusion matrix that shows the comparison's results of the devices in the case of Learning with W=60, Testing with W=80, getting a diagonal-rate of 14/14}
         \label{heat-map-20diff}
     \end{subfigure}
     \hfill
     \begin{subfigure}[t]{0.45\textwidth}
         \centering
         \captionsetup{justification=centering}
         \includegraphics[width=\textwidth]{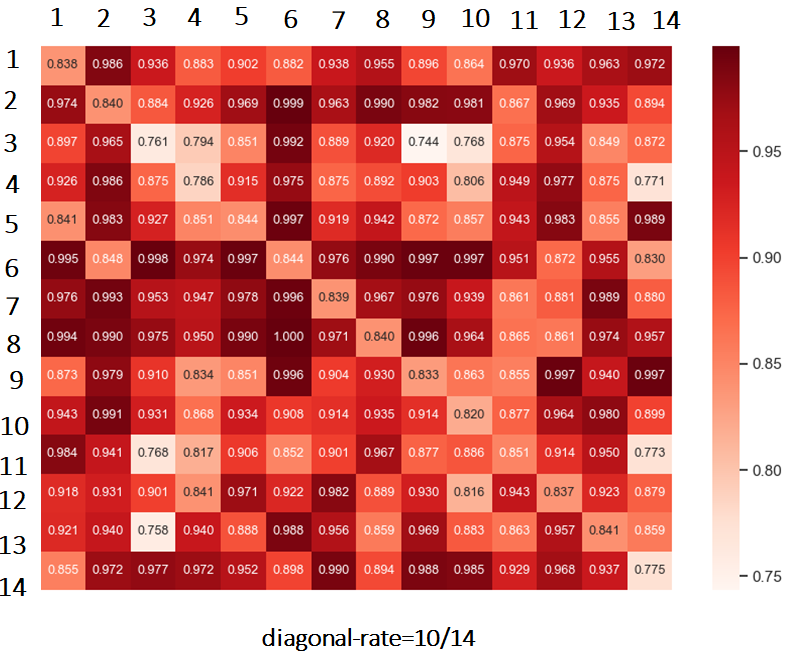}
         \caption{Confusion matrix - Learning with W=40, Testing with W=80}
         \Description{Confusion matrix that shows the comparison's results of the devices in the case of Learning with W=40, Testing with W=80, getting a diagonal-rate of 10/14}
         \label{heat-map-40diff}
     \end{subfigure}
        \caption{Detecting dominating device with different learnt and tested W}
        \Description{Detecting dominating device with different learnt and tested W}
        \label{Detecting the dominating device with different learnt and tested W}
\end{figure*}
\section{Estimating the shaping and padding parameters}\label{Identifying the shaping and padding parameters}
\subsection{Identifying Padding Parameter $W$}\label{Identifying Parameter $W$ Regarding the STP}
As mentioned in Section \ref{Padding And Shaping}, we model a case which STP pads the traffic randomly, using the parameter $W$. In order for the adversary to model the devices and to identify them accurately, ideally it needs to know $W$. 
We evaluated what happens if there is a difference between the actual $W$ and the value $W$ estimated by the adversary. Figure ~\ref{Detecting the dominating device with different learnt and tested W} repeats the experiment of Section \ref{Identifying Active Dominating Device} when the adversary estimates $W$ incorrectly. Figure ~\ref{heat-map-20diff} shows that the adversary can still identify correctly the dominating active devices when under-estimating W by 20 bytes. Figure ~\ref{heat-map-40diff} has a diagonal-rate of 10/14 and shows that a larger under-estimation (of 40) yields to incorrect classifications. 

Now, in order to estimate $W$ with reasonable accuracy, the adversary does the following: whenever there's indication that only one device is active (e.g., a dominating device), the adversary compares the recorded trace with the models of all of the devices, for all the different values of $W$ it modeled. As discussed earlier, we know that taking the modeled devices with $W$ at most 20 bytes away from the actual value would yield a true classification of that device's identity with high likelihood. Therefore it's sufficient for us to find $W$ which is at most 20 bytes off from the actual $W$. 

\begin{figure}
     \centering
     \captionsetup{justification=centering}
     \begin{subfigure}[t]{0.3\linewidth}
         \centering
         \captionsetup{justification=centering}
         \includegraphics[width=0.3\linewidth]{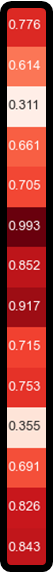}
         \caption{Learning with W=50, Testing with W=80}
         \Description{Column's values that show the comparison's results of the devices in the case of Learning with W=50, Testing with W=80}
         \label{Confusion matrix - Learning with W=50, Testing with W=80}
     \end{subfigure}
     \hfill
     \begin{subfigure}[t]{0.3\linewidth}
         \centering
         \captionsetup{justification=centering}
         \includegraphics[width=0.3\linewidth]{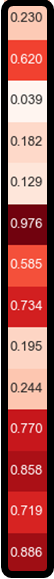}
         \caption{Learning with W=90, Testing with W=80}
         \Description{Column's values that show the comparison's results of the devices in the case of Learning with W=90, Testing with W=80}
         \label{Confusion matrix - Learning with W=90, Testing with W=80}
     \end{subfigure}
     \hfill
     \begin{subfigure}[t]{0.3\linewidth}
         \centering
         \captionsetup{justification=centering}
         \includegraphics[width=0.3\linewidth]{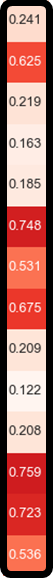}
         \caption{Learning with W=130, Testing with W=80}
         \Description{Column's values that show the comparison's results of the devices in the case of Learning with W=130, Testing with W=80}
         \label{Confusion matrix - Learning with W=130, Testing with W=80}
     \end{subfigure}
        \caption{Detecting W - Example when device 3 is the dominating device}
        \Description{Example that show the comparison's results of the devices for each checked value of W, when device 3 is the dominating device }
        \label{Detecting W}
\end{figure}

To evaluate the adversary's power, we simulate the padded traffic of every device for multiple possible values of the parameter $W$ (testing with jumps of 40 bytes to aim for $\pm$ 20 accuracy). Then we model the learnt devices for the different values of $W$. Whenever there is an active dominating device whose identity is unknown to the adversary, it compares it to the different models for each value of $W$ we checked.
Figure ~\ref{Detecting W} shows an example of the derived models' columns comparison (using cosine distance) to the tested device, when the true value is $W=80$ and we simulate with $W=50,90,130$ for one possible device (in this case, $W=90$ has the best estimate of $W$ assuming device~3 is dominant). 

The smallest value's row index in each column indicates which device is estimated to be active when we simulate with the associated value of $W$, and the row index of the minimum of all minima indicate which device is the best candidate out of all columns. The column associated with this minimum has the closest value of $W$ to the real value used. In Figure ~\ref{Detecting W} we choose $W=90$, because its column has the minimum value 0.039. Since the jump between tested $W$ values is 40, we conclude that the true value is in the values $90\pm 20$.

\begin{figure}[t]
         \centering
         \captionsetup{justification=centering}
         \includegraphics[width=\linewidth]{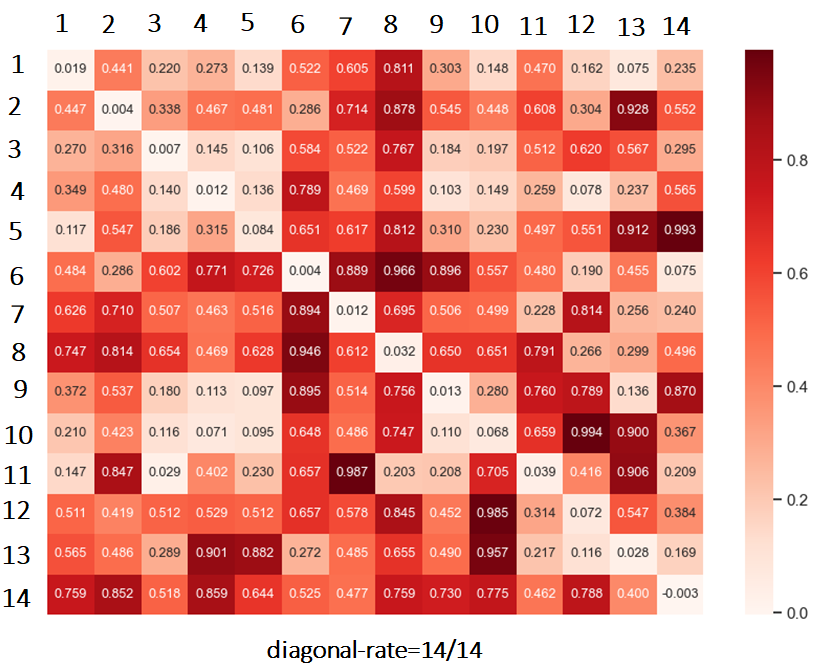}
         \caption{Confusion matrix - \\
         Learning with q=0.5, Testing with q=0.1}
         \Description{Confusion matrix that shows the comparison's results of the devices in the case of Learning with q=0.5 and Testing with q=0.1, getting a diagonal-rate of 14/14}
         \label{Confusion matrix - Learning with q=0.5, Testing with q=0.1}
\end{figure}

\begin{figure}[t]
         \centering
         \captionsetup{justification=centering}
         \includegraphics[width=\linewidth]{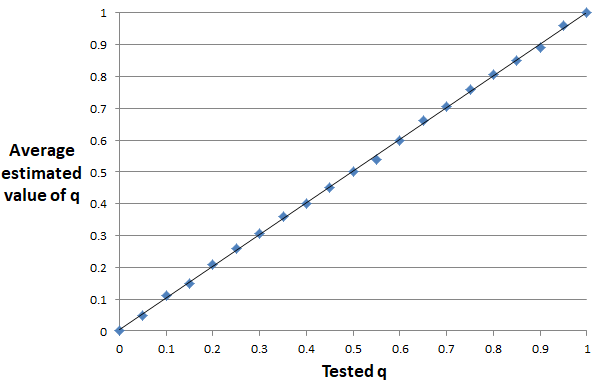}
         \caption{Estimating the value of tested q over all the devices}
         \Description{Plot that show the average estimated value of q as function of the real value of q, demonstrating that they are always very close}
         \label{Estimating the value of tested q over all the devices}
\end{figure}

\subsection{Identifying Shaping Parameter $q$}\label{Identifying Parameter $q$ Regarding the STP}
Next the adversary would like to estimate the shaping probability $q$ used by the victim.
We again assume that a single (dominating) device is active, build simulated models for multiple values of $q$ and compare them to the actual traffic with the unknown $q$. Figure ~\ref{Confusion matrix - Learning with q=0.5, Testing with q=0.1} demonstrates that even when we model with value of $q$ that is very different from the real value used, we still achieve diagonal-rate of 14/14. 

Nonetheless, we still wish to get a reasonable estimate of $q$. We observe that $q$ directly affects the traffic volume: the higher $q$ is, the higher the pps rate is.

Therefore we check the traffic volume emitted by the dominant device for different values of $q$, and set thresholds in the mid-points between them (similar to what we did for the Number Of Active Devices Classifier in Section ~\ref{Identifying Number Of Active Devices}). 
By checking the actual traffic volume recorded in real-time and comparing it to the determined thresholds we can estimate $q$.

To evaluate this estimation, we checked values of $q$ that range between 0 and 1 with jump of $0.05$.
Figure ~\ref{Estimating the value of tested q over all the devices} shows the average estimated $q$ (over all the devices) as a function of the true $q$. The figure shows that in all cases the estimation is very close to the true value:
we precisely estimate the value of $q$ with accuracy of 91\%, and in all cases we estimate the value of $q$ to be at most $0.1$ off from the actual value.

\section{Detecting Periods with Real Traffic}\label{Detecting Periods with Real Traffic}

Suppose now that the external adversary, that knows that shaping and padding are being employed on the middlebox home gateway, wants to distinguish between real traffic and cover (false) traffic injected by the STP algorithm.
As before, we focus on using the STP with $q=0.1, W=80, T=1sec, R=100pps$.

Recall that STP breaks up time into \emph{periods} of length T. If the device is active in period $j$ then STP shapes the traffic so the packets are emitted uniformly, at a rate of $R$ pps, with equal inter-packet-arrival times (and padded lengths). If the device is not active in period $j$, STP randomly (with probability $q$) emits exactly $R/T$ equally-spaced cover packets in period $j$. 

We assume that the adversary knows $T$, but is not synchronized with the gateway. Thus, the adversary also splits time into time units of length $T$: for ease of exposition we call the adversary's time units ``time windows'' to distinguish them from the STP's ``periods''. A time window overlaps two periods.
Notice that time windows can be empty when both STP periods it overlaps contain no real traffic and STP randomly decided not to inject cover traffic in either.

Each non-empty window might contain real traffic, or it may contain only cover traffic. The adversary's goal is to label the non-empty windows, as it splits them, into two classes: ``only cover traffic'' and ``real traffic''.

Dividing the training-trace into time windows splits each period into two windows, so we get more non-empty time windows than non-empty periods, possible twice as many. With our devices and the STP parameters as  before we observed that 72\% of the non-empty periods contain real traffic, while 53\% of non-empty windows contain real traffic. Therefore, the partition to windows instead of periods yields a more balanced training set, where about half of the windows are labeled as ``real traffic'' and the other half as ``only cover traffic''.

For the purpose of classifying the non-empty windows between the two classes, we create a feature vector for each time window. Since STP shapes the traffic, each time window effectively has $T/R$ ``slots'' at which a packet is either emitted or not. Thus for a time window starting at time $t_{start}$ we define a feature vector of size of $T/R$ where the $i$th feature ($0 \leq i\leq  T/R  -1$)  represents the size of the packet in the $i$th slot (at time $t_{start}+ i*R$), or $0$ if there is no packet in that slot. Note that, unlike the packet-sizes distributions we used before, this feature is sensitive to packet-size order, reminiscent of the DFA model \cite{anomaly-dfa}.

The adversary uses the labeled windows as a classified training set for both classes of windows.
Then the adversary divides the test trace into time windows in the same way and labels any non-empty tested window using its training set.

We tried different machine learning algorithms such as Random Forest (RF) and Support Vector Machine (SVM) but found that the most effective algorithm is the KNN algorithm. 
To set the value of $k$, we applied 10-fold cross validation on the training set, each time with a different value of $k$, for $k$ between 1 to 150, and chose the best value $k=119$ that results in the highest accuracy of the model.

To evaluate the model, we tested it on a different trace and repeated this process for 10 times, each time with a different subset of active devices. The results in Table \ref{detecting real periods} demonstrate that we managed to obtain high measures of Precision and Recall (above 77\%) and accuracy of 81\%. 

\begin{table}[t]
\centering
 \captionsetup{justification=centering}
\caption{Classifying The time windows Using KNN with k=119}
\begin{tabular}{|P{2.3cm}|P{2.3cm}|P{2.3cm}|}
\hline

Recall

(\%)&Precision

(\%)&Accuracy

(\%)\\\hline
78 & 77 & 81  \\\hline
\end{tabular}
\label{detecting real periods}
\end{table}

\section{Conclusion}\label{Future Work}

In this paper we demonstrated that the packet-sizes distribution can still be used to successfully fingerprint the active IoT devices when shaping and padding are used---as long as the adversary is aware that these mitigations are deployed, and even if the values of the padding and shaping parameters are unknown. The main tool we used in our analysis is the full distribution of packet-sizes---as opposed to commonly used statistics such as mean and variance. We further showed how an external adversary who only sees the padded and shaped traffic as aggregated and hidden behind a NAT middlebox can accurately identify the subset of active devices with Recall and Precision of at least 96\%. We also showed that the adversary can distinguish time windows containing only bogus cover packets from windows with real device activity, at a granularity of $1sec$ time windows, with 81\% accuracy. Using similar methodology, but now on the defender's side, we were also able to detect anomalous activities in IoT traffic due to the Mirai worm.

\bibliographystyle{ACM-Reference-Format}
\bibliography{References.bib}


\begin{thebibliography}{44}


\ifx \showCODEN    \undefined \def \showCODEN     #1{\unskip}     \fi
\ifx \showDOI      \undefined \def \showDOI       #1{#1}\fi
\ifx \showISBNx    \undefined \def \showISBNx     #1{\unskip}     \fi
\ifx \showISBNxiii \undefined \def \showISBNxiii  #1{\unskip}     \fi
\ifx \showISSN     \undefined \def \showISSN      #1{\unskip}     \fi
\ifx \showLCCN     \undefined \def \showLCCN      #1{\unskip}     \fi
\ifx \shownote     \undefined \def \shownote      #1{#1}          \fi
\ifx \showarticletitle \undefined \def \showarticletitle #1{#1}   \fi
\ifx \showURL      \undefined \def \showURL       {\relax}        \fi
\providecommand\bibfield[2]{#2}
\providecommand\bibinfo[2]{#2}
\providecommand\natexlab[1]{#1}
\providecommand\showeprint[2][]{arXiv:#2}

\bibitem[\protect\citeauthoryear{Acar, Fereidooni, Abera, Sikder, Miettinen,
  Aksu, Conti, Sadeghi, and Uluagac}{Acar et~al\mbox{.}}{2020}]%
        {Peek-a-boo}
\bibfield{author}{\bibinfo{person}{Abbas Acar}, \bibinfo{person}{Hossein
  Fereidooni}, \bibinfo{person}{Tigist Abera}, \bibinfo{person}{Amit~Kumar
  Sikder}, \bibinfo{person}{Markus Miettinen}, \bibinfo{person}{Hidayet Aksu},
  \bibinfo{person}{Mauro Conti}, \bibinfo{person}{Ahmad-Reza Sadeghi}, {and}
  \bibinfo{person}{Selcuk Uluagac}.} \bibinfo{year}{2020}\natexlab{}.
\newblock \showarticletitle{Peek-a-Boo: I See Your Smart Home Activities, Even
  Encrypted!}. In \bibinfo{booktitle}{\emph{Proceedings of the 13th ACM
  Conference on Security and Privacy in Wireless and Mobile Networks}}
  \emph{(\bibinfo{series}{WiSec '20})}. \bibinfo{publisher}{Association for
  Computing Machinery}, \bibinfo{address}{New York, NY, USA},
  \bibinfo{pages}{207–218}.
\newblock
\showISBNx{9781450380065}
\urldef\tempurl%
\url{https://doi.org/10.1145/3395351.3399421}
\showDOI{\tempurl}


\bibitem[\protect\citeauthoryear{Alghushairy, Alsini, Soule, and
  Ma}{Alghushairy et~al\mbox{.}}{2021}]%
        {lof-2}
\bibfield{author}{\bibinfo{person}{Omar Alghushairy}, \bibinfo{person}{Raed
  Alsini}, \bibinfo{person}{Terence Soule}, {and} \bibinfo{person}{Xiaogang
  Ma}.} \bibinfo{year}{2021}\natexlab{}.
\newblock \showarticletitle{A Review of Local Outlier Factor Algorithms for
  Outlier Detection in Big Data Streams}.
\newblock \bibinfo{journal}{\emph{Big Data and Cognitive Computing}}
  \bibinfo{volume}{5}, \bibinfo{number}{1} (\bibinfo{year}{2021}).
\newblock
\showISSN{2504-2289}
\urldef\tempurl%
\url{https://doi.org/10.3390/bdcc5010001}
\showDOI{\tempurl}


\bibitem[\protect\citeauthoryear{Alshehri, Granley, and Yue}{Alshehri
  et~al\mbox{.}}{2020}]%
        {Random-Padding}
\bibfield{author}{\bibinfo{person}{Ahmed Alshehri}, \bibinfo{person}{Jacob
  Granley}, {and} \bibinfo{person}{Chuan Yue}.}
  \bibinfo{year}{2020}\natexlab{}.
\newblock \showarticletitle{Attacking and Protecting Tunneled Traffic of Smart
  Home Devices}. In \bibinfo{booktitle}{\emph{Proceedings of the Tenth ACM
  Conference on Data and Application Security and Privacy}}
  \emph{(\bibinfo{series}{CODASPY '20})}. \bibinfo{publisher}{Association for
  Computing Machinery}, \bibinfo{address}{New York, NY, USA},
  \bibinfo{pages}{259–270}.
\newblock
\showISBNx{9781450371070}
\urldef\tempurl%
\url{https://doi.org/10.1145/3374664.3375723}
\showDOI{\tempurl}


\bibitem[\protect\citeauthoryear{Antonakakis, April, Bailey, Bernhard,
  Bursztein, Cochran, Durumeric, Halderman, Invernizzi, Kallitsis, Kumar,
  Lever, Ma, Mason, Menscher, Seaman, Sullivan, Thomas, and Zhou}{Antonakakis
  et~al\mbox{.}}{2017}]%
        {mirai-3}
\bibfield{author}{\bibinfo{person}{Manos Antonakakis}, \bibinfo{person}{Tim
  April}, \bibinfo{person}{Michael Bailey}, \bibinfo{person}{Matt Bernhard},
  \bibinfo{person}{Elie Bursztein}, \bibinfo{person}{Jaime Cochran},
  \bibinfo{person}{Zakir Durumeric}, \bibinfo{person}{J.~Alex Halderman},
  \bibinfo{person}{Luca Invernizzi}, \bibinfo{person}{Michalis Kallitsis},
  \bibinfo{person}{Deepak Kumar}, \bibinfo{person}{Chaz Lever},
  \bibinfo{person}{Zane Ma}, \bibinfo{person}{Joshua Mason},
  \bibinfo{person}{Damian Menscher}, \bibinfo{person}{Chad Seaman},
  \bibinfo{person}{Nick Sullivan}, \bibinfo{person}{Kurt Thomas}, {and}
  \bibinfo{person}{Yi Zhou}.} \bibinfo{year}{2017}\natexlab{}.
\newblock \showarticletitle{Understanding the Mirai Botnet}. In
  \bibinfo{booktitle}{\emph{26th {USENIX} Security Symposium ({USENIX} Security
  17)}}. \bibinfo{publisher}{{USENIX} Association},
  \bibinfo{address}{Vancouver, BC}, \bibinfo{pages}{1093--1110}.
\newblock
\showISBNx{978-1-931971-40-9}
\urldef\tempurl%
\url{https://www.usenix.org/conference/usenixsecurity17/technical-sessions/presentation/antonakakis}
\showURL{%
\tempurl}


\bibitem[\protect\citeauthoryear{Apthorpe, Reisman, and Feamster}{Apthorpe
  et~al\mbox{.}}{2017a}]%
        {apthorpe-1}
\bibfield{author}{\bibinfo{person}{Apthorpe}, \bibinfo{person}{Reisman}, {and}
  \bibinfo{person}{Feamster}.} \bibinfo{year}{2017}\natexlab{a}.
\newblock \showarticletitle{Closing the Blinds: Four Strategies for Protecting
  Smart Home Privacy from Network Observers}. In \bibinfo{booktitle}{\emph{IEEE
  Workshop on Technology and Consumer Protection (ConPro)}}.
  \bibinfo{address}{San Francisco, CA}, \bibinfo{pages}{1--6}.
\newblock
\urldef\tempurl%
\url{https://www.ieee-security.org/TC/SPW2017/ConPro/papers/apthorpe-conpro17.pdf}
\showURL{%
\tempurl}


\bibitem[\protect\citeauthoryear{Apthorpe, Huang, Reisman, Narayanan, and
  Feamster}{Apthorpe et~al\mbox{.}}{2019}]%
        {STP}
\bibfield{author}{\bibinfo{person}{Noah Apthorpe}, \bibinfo{person}{Danny
  Huang}, \bibinfo{person}{Dillon Reisman}, \bibinfo{person}{Arvind Narayanan},
  {and} \bibinfo{person}{Nick Feamster}.} \bibinfo{year}{2019}\natexlab{}.
\newblock \showarticletitle{Keeping the Smart Home Private with Smart(er) IoT
  Traffic Shaping}.
\newblock \bibinfo{journal}{\emph{Proceedings on Privacy Enhancing
  Technologies}}  \bibinfo{volume}{2019}, \bibinfo{pages}{128--148}.
\newblock
\urldef\tempurl%
\url{https://doi.org/10.2478/popets-2019-0040}
\showDOI{\tempurl}


\bibitem[\protect\citeauthoryear{Apthorpe, Reisman, Sundaresan, Narayanan, and
  Feamster}{Apthorpe et~al\mbox{.}}{2017b}]%
        {apthorpe-3}
\bibfield{author}{\bibinfo{person}{Noah Apthorpe}, \bibinfo{person}{Dillon
  Reisman}, \bibinfo{person}{Srikanth Sundaresan}, \bibinfo{person}{Arvind
  Narayanan}, {and} \bibinfo{person}{Nick Feamster}.}
  \bibinfo{year}{2017}\natexlab{b}.
\newblock \bibinfo{booktitle}{\emph{Spying on the Smart Home: Privacy Attacks
  and Defenses on Encrypted IoT Traffic}}.
\newblock \bibinfo{type}{{T}echnical {R}eport} arXiv:1708.05044 [cs.CR].
\newblock


\bibitem[\protect\citeauthoryear{Apthorpe, Reissman, and Feamster}{Apthorpe
  et~al\mbox{.}}{2016}]%
        {apthorpe-2}
\bibfield{author}{\bibinfo{person}{Noah Apthorpe}, \bibinfo{person}{Dillon
  Reissman}, {and} \bibinfo{person}{Nick Feamster}.}
  \bibinfo{year}{2016}\natexlab{}.
\newblock \showarticletitle{A Smart Home is No Castle: Privacy Vulnerabilities
  of Encrypted IoT Traffic}. In \bibinfo{booktitle}{\emph{Workshop on Data and
  Algorithmic Transparency (DAT)}}. \bibinfo{address}{New York, NY},
  \bibinfo{pages}{1--6}.
\newblock
\urldef\tempurl%
\url{http://datworkshop.org/papers/dat16-final37.pdf}
\showURL{%
\tempurl}


\bibitem[\protect\citeauthoryear{Bezawada, Bachani, Peterson, Shirazi, Ray, and
  Ray}{Bezawada et~al\mbox{.}}{2018}]%
        {IoTSense}
\bibfield{author}{\bibinfo{person}{Bruhadeshwar Bezawada},
  \bibinfo{person}{Maalvika Bachani}, \bibinfo{person}{Jordan Peterson},
  \bibinfo{person}{Hossein Shirazi}, \bibinfo{person}{Indrakshi Ray}, {and}
  \bibinfo{person}{Indrajit Ray}.} \bibinfo{year}{2018}\natexlab{}.
\newblock \showarticletitle{IoTSense: Behavioral Fingerprinting of IoT
  Devices}.
\newblock \bibinfo{journal}{\emph{arXiv preprint arXiv:1804.03852}}
  (\bibinfo{date}{04} \bibinfo{year}{2018}).
\newblock


\bibitem[\protect\citeauthoryear{Breunig, Kriegel, Ng, and Sander}{Breunig
  et~al\mbox{.}}{2000}]%
        {lof-1}
\bibfield{author}{\bibinfo{person}{Markus Breunig}, \bibinfo{person}{Hans-Peter
  Kriegel}, \bibinfo{person}{Raymond Ng}, {and} \bibinfo{person}{Joerg
  Sander}.} \bibinfo{year}{2000}\natexlab{}.
\newblock \showarticletitle{LOF: Identifying Density-Based Local Outliers.}
\newblock \bibinfo{journal}{\emph{ACM Sigmod Record}}  \bibinfo{volume}{29},
  \bibinfo{pages}{93--104}.
\newblock
\urldef\tempurl%
\url{https://doi.org/10.1145/342009.335388}
\showDOI{\tempurl}


\bibitem[\protect\citeauthoryear{Connor, Cardillo, Moss, and Rabitti}{Connor
  et~al\mbox{.}}{2013}]%
        {kl-1}
\bibfield{author}{\bibinfo{person}{Richard Connor},
  \bibinfo{person}{Franco~Alberto Cardillo}, \bibinfo{person}{Robert Moss},
  {and} \bibinfo{person}{Fausto Rabitti}.} \bibinfo{year}{2013}\natexlab{}.
\newblock \showarticletitle{Evaluation of Jensen-Shannon Distance over Sparse
  Data}. In \bibinfo{booktitle}{\emph{Similarity Search and Applications}},
  \bibfield{editor}{\bibinfo{person}{Nieves Brisaboa}, \bibinfo{person}{Oscar
  Pedreira}, {and} \bibinfo{person}{Pavel Zezula}} (Eds.).
  \bibinfo{publisher}{Springer Berlin Heidelberg}, \bibinfo{address}{Berlin,
  Heidelberg}, \bibinfo{pages}{163--168}.
\newblock


\bibitem[\protect\citeauthoryear{Conti, Mancini, Spolaor, and Verde}{Conti
  et~al\mbox{.}}{2015}]%
        {Analyzing-Android-Encrypted}
\bibfield{author}{\bibinfo{person}{Mauro Conti}, \bibinfo{person}{Luigi
  Mancini}, \bibinfo{person}{Riccardo Spolaor}, {and} \bibinfo{person}{Nino
  Verde}.} \bibinfo{year}{2015}\natexlab{}.
\newblock \showarticletitle{Analyzing Android Encrypted Network Traffic to
  Identify User Actions}.
\newblock \bibinfo{journal}{\emph{IEEE Transactions on Information Forensics
  and Security}}  \bibinfo{volume}{11} (\bibinfo{date}{01}
  \bibinfo{year}{2015}), \bibinfo{pages}{1--1}.
\newblock
\urldef\tempurl%
\url{https://doi.org/10.1109/TIFS.2015.2478741}
\showDOI{\tempurl}


\bibitem[\protect\citeauthoryear{Dong, Li, Tang, Chen, Sun, and Zhang}{Dong
  et~al\mbox{.}}{2020}]%
        {deep-2}
\bibfield{author}{\bibinfo{person}{Shuaike Dong}, \bibinfo{person}{Zhou Li},
  \bibinfo{person}{Di Tang}, \bibinfo{person}{Jiongyi Chen},
  \bibinfo{person}{Menghan Sun}, {and} \bibinfo{person}{Kehuan Zhang}.}
  \bibinfo{year}{2020}\natexlab{}.
\newblock \showarticletitle{Your Smart Home Can't Keep a Secret: Towards
  Automated Fingerprinting of IoT Traffic}. In
  \bibinfo{booktitle}{\emph{Proceedings of the 15th ACM Asia Conference on
  Computer and Communications Security}} (Taipei, Taiwan)
  \emph{(\bibinfo{series}{ASIA CCS '20})}. \bibinfo{publisher}{Association for
  Computing Machinery}, \bibinfo{address}{New York, NY, USA},
  \bibinfo{pages}{47–59}.
\newblock
\showISBNx{9781450367509}
\urldef\tempurl%
\url{https://doi.org/10.1145/3320269.3384732}
\showDOI{\tempurl}


\bibitem[\protect\citeauthoryear{Dyer, Coull, Ristenpart, and Shrimpton}{Dyer
  et~al\mbox{.}}{2012}]%
        {padding-1}
\bibfield{author}{\bibinfo{person}{Kevin~P. Dyer}, \bibinfo{person}{Scott~E.
  Coull}, \bibinfo{person}{Thomas Ristenpart}, {and} \bibinfo{person}{Thomas
  Shrimpton}.} \bibinfo{year}{2012}\natexlab{}.
\newblock \showarticletitle{Peek-a-Boo, I Still See You: Why Efficient Traffic
  Analysis Countermeasures Fail}. In \bibinfo{booktitle}{\emph{2012 IEEE
  Symposium on Security and Privacy}}. \bibinfo{pages}{332--346}.
\newblock
\urldef\tempurl%
\url{https://doi.org/10.1109/SP.2012.28}
\showDOI{\tempurl}


\bibitem[\protect\citeauthoryear{Dziubinski and Bandai}{Dziubinski and
  Bandai}{2021}]%
        {shaping-1}
\bibfield{author}{\bibinfo{person}{Kiana Dziubinski} {and}
  \bibinfo{person}{Masaki Bandai}.} \bibinfo{year}{2021}\natexlab{}.
\newblock \showarticletitle{Bandwidth Efficient IoT Traffic Shaping Technique
  for Protecting Smart Home Privacy from Data Breaches in Wireless LAN}.
\newblock \bibinfo{journal}{\emph{IEICE Transactions on Communications}}
  (\bibinfo{date}{02} \bibinfo{year}{2021}).
\newblock
\urldef\tempurl%
\url{https://doi.org/10.1587/transcom.2020EBP3182}
\showDOI{\tempurl}


\bibitem[\protect\citeauthoryear{Garcia, Alcaniz, González-Vidal, Bernabe,
  Rivera, and Skarmeta}{Garcia et~al\mbox{.}}{2021}]%
        {anomaly-deep-2}
\bibfield{author}{\bibinfo{person}{Norberto Garcia}, \bibinfo{person}{Tomas
  Alcaniz}, \bibinfo{person}{Aurora González-Vidal},
  \bibinfo{person}{Jorge~Bernal Bernabe}, \bibinfo{person}{Diego Rivera}, {and}
  \bibinfo{person}{Antonio Skarmeta}.} \bibinfo{year}{2021}\natexlab{}.
\newblock \showarticletitle{Distributed real-time SlowDoS attacks detection
  over encrypted traffic using Artificial Intelligence}.
\newblock \bibinfo{journal}{\emph{Journal of Network and Computer
  Applications}}  \bibinfo{volume}{173} (\bibinfo{year}{2021}),
  \bibinfo{pages}{102871}.
\newblock
\showISSN{1084-8045}
\urldef\tempurl%
\url{https://doi.org/10.1016/j.jnca.2020.102871}
\showDOI{\tempurl}


\bibitem[\protect\citeauthoryear{Goldenberg and Wool}{Goldenberg and
  Wool}{2013}]%
        {gw13}
\bibfield{author}{\bibinfo{person}{N. Goldenberg} {and} \bibinfo{person}{A.
  Wool}.} \bibinfo{year}{2013}\natexlab{}.
\newblock \showarticletitle{Accurate Modeling of {Modbus/TCP} for Intrusion
  Detection in {SCADA} Systems}.
\newblock \bibinfo{journal}{\emph{International Journal of Critical
  Infrastructure Protection}}  \bibinfo{volume}{6} (\bibinfo{year}{2013}),
  \bibinfo{pages}{63--75}.
\newblock
\urldef\tempurl%
\url{http://dx.doi.org/10.1016/j.ijcip.2013.05.001}
\showURL{%
\tempurl}


\bibitem[\protect\citeauthoryear{Haji and Ameen}{Haji and Ameen}{2021}]%
        {anomaly-ml-1}
\bibfield{author}{\bibinfo{person}{Saad Haji} {and} \bibinfo{person}{Siddeeq
  Ameen}.} \bibinfo{year}{2021}\natexlab{}.
\newblock \showarticletitle{Attack and Anomaly Detection in IoT Networks using
  Machine Learning Techniques: A Review}.
\newblock \bibinfo{journal}{\emph{Asian Journal of Research in Computer
  Science}} (\bibinfo{date}{06} \bibinfo{year}{2021}), \bibinfo{pages}{30--46}.
\newblock
\urldef\tempurl%
\url{https://doi.org/10.9734/ajrcos/2021/v9i230218}
\showDOI{\tempurl}


\bibitem[\protect\citeauthoryear{Holst}{Holst}{2021}]%
        {how-many-iot-2}
\bibfield{author}{\bibinfo{person}{Arne Holst}.}
  \bibinfo{year}{2021}\natexlab{}.
\newblock \bibinfo{booktitle}{\emph{Number of Internet of Things (IoT)
  connected devices worldwide from 2019 to 2030}}.
\newblock
\urldef\tempurl%
\url{https://www.statista.com/statistics/1183457/iot-connected-devices-worldwide/}
\showURL{%
\tempurl}


\bibitem[\protect\citeauthoryear{Jerkins}{Jerkins}{2017}]%
        {Mirai-Ref1}
\bibfield{author}{\bibinfo{person}{James Jerkins}.}
  \bibinfo{year}{2017}\natexlab{}.
\newblock \showarticletitle{Motivating a market or regulatory solution to IoT
  insecurity with the Mirai botnet code}. In \bibinfo{booktitle}{\emph{Proc.
  IEEE 7th computing and communication workshop and conference (CCWC)}}.
  \bibinfo{pages}{1--5}.
\newblock
\urldef\tempurl%
\url{https://doi.org/10.1109/CCWC.2017.7868464}
\showDOI{\tempurl}


\bibitem[\protect\citeauthoryear{Jonsdottir, Wood, and Doshi}{Jonsdottir
  et~al\mbox{.}}{2017}]%
        {anomaly-ml-3}
\bibfield{author}{\bibinfo{person}{Gudrun Jonsdottir}, \bibinfo{person}{Daniel
  Wood}, {and} \bibinfo{person}{Rohan Doshi}.} \bibinfo{year}{2017}\natexlab{}.
\newblock \showarticletitle{IoT network monitor}. In
  \bibinfo{booktitle}{\emph{2017 IEEE MIT Undergraduate Research Technology
  Conference (URTC)}}. \bibinfo{pages}{1--5}.
\newblock
\urldef\tempurl%
\url{https://doi.org/10.1109/URTC.2017.8284179}
\showDOI{\tempurl}


\bibitem[\protect\citeauthoryear{Junges, François, and Festor}{Junges
  et~al\mbox{.}}{2019}]%
        {Passive-Inference}
\bibfield{author}{\bibinfo{person}{Pierre-Marie Junges}, \bibinfo{person}{J.
  François}, {and} \bibinfo{person}{O. Festor}.}
  \bibinfo{year}{2019}\natexlab{}.
\newblock \showarticletitle{Passive Inference of User Actions through IoT
  Gateway Encrypted Traffic Analysis}.
\newblock \bibinfo{journal}{\emph{2019 IFIP/IEEE Symposium on Integrated
  Network and Service Management (IM)}} (\bibinfo{year}{2019}),
  \bibinfo{pages}{7--12}.
\newblock


\bibitem[\protect\citeauthoryear{Lam and Abbas}{Lam and Abbas}{2020}]%
        {anomaly-deep-1}
\bibfield{author}{\bibinfo{person}{Jordan Lam} {and} \bibinfo{person}{Robert
  Abbas}.} \bibinfo{year}{2020}\natexlab{}.
\newblock \showarticletitle{Machine Learning based Anomaly Detection for 5G
  Networks}.
\newblock \bibinfo{journal}{\emph{ArXiv}}  \bibinfo{volume}{abs/2003.03474}
  (\bibinfo{year}{2020}).
\newblock


\bibitem[\protect\citeauthoryear{learn developers}{learn developers}{2020}]%
        {sklearn}
\bibfield{author}{\bibinfo{person}{Scikit learn developers}.}
  \bibinfo{year}{2007 - 2020}\natexlab{}.
\newblock \bibinfo{booktitle}{\emph{Unsupervised Outlier Detection using Local
  Outlier Factor (LOF)- Scikit-learn library}}.
\newblock
\urldef\tempurl%
\url{https://scikit-learn.org/stable/modules/generated/sklearn.neighbors.LocalOutlierFactor.html}
\showURL{%
\tempurl}


\bibitem[\protect\citeauthoryear{Lee}{Lee}{1999}]%
        {cosine/kl}
\bibfield{author}{\bibinfo{person}{Lillian Lee}.}
  \bibinfo{year}{1999}\natexlab{}.
\newblock \showarticletitle{Measures of Distributional Similarity}. In
  \bibinfo{booktitle}{\emph{Proceedings of the 37th Annual Meeting of the
  Association for Computational Linguistics on Computational Linguistics}}
  (College Park, Maryland) \emph{(\bibinfo{series}{ACL '99})}.
  \bibinfo{publisher}{Association for Computational Linguistics},
  \bibinfo{address}{USA}, \bibinfo{pages}{25–32}.
\newblock
\showISBNx{1558606093}
\urldef\tempurl%
\url{https://doi.org/10.3115/1034678.1034693}
\showDOI{\tempurl}


\bibitem[\protect\citeauthoryear{Liao and Xu}{Liao and Xu}{2015}]%
        {cosine-1}
\bibfield{author}{\bibinfo{person}{Huchang Liao} {and} \bibinfo{person}{Zeshui
  Xu}.} \bibinfo{year}{2015}\natexlab{}.
\newblock \showarticletitle{Approaches to manage hesitant fuzzy linguistic
  information based on the cosine distance and similarity measures for HFLTSs
  and their application in qualitative decision making}.
\newblock \bibinfo{journal}{\emph{Expert Systems with Applications}}
  \bibinfo{volume}{42}, \bibinfo{number}{12} (\bibinfo{year}{2015}),
  \bibinfo{pages}{5328--5336}.
\newblock
\showISSN{0957-4174}
\urldef\tempurl%
\url{https://doi.org/10.1016/j.eswa.2015.02.017}
\showDOI{\tempurl}


\bibitem[\protect\citeauthoryear{McDermott, Haynes, and Petrovksi}{McDermott
  et~al\mbox{.}}{2018}]%
        {Attack-Vectors}
\bibfield{author}{\bibinfo{person}{Christopher McDermott},
  \bibinfo{person}{William Haynes}, {and} \bibinfo{person}{Andrei Petrovksi}.}
  \bibinfo{year}{2018}\natexlab{}.
\newblock \showarticletitle{Threat Detection and Analysis in the Internet of
  Things using Deep Packet Inspection}.
\newblock \bibinfo{journal}{\emph{International Journal on Cyber Situational
  Awareness}}  \bibinfo{volume}{4} (\bibinfo{date}{12} \bibinfo{year}{2018}),
  \bibinfo{pages}{61--83}.
\newblock
\urldef\tempurl%
\url{https://doi.org/10.22619/IJCSA.2018.100120}
\showDOI{\tempurl}


\bibitem[\protect\citeauthoryear{Meidan, Bohadana, Shabtai, Guarnizo, Ochoa,
  Tippenhauer, and Elovici}{Meidan et~al\mbox{.}}{2017}]%
        {Beer-Sheva}
\bibfield{author}{\bibinfo{person}{Yair Meidan}, \bibinfo{person}{Michael
  Bohadana}, \bibinfo{person}{Asaf Shabtai}, \bibinfo{person}{Juan~David
  Guarnizo}, \bibinfo{person}{Mart\'{\i}n Ochoa}, \bibinfo{person}{Nils~Ole
  Tippenhauer}, {and} \bibinfo{person}{Yuval Elovici}.}
  \bibinfo{year}{2017}\natexlab{}.
\newblock \showarticletitle{ProfilIoT: A Machine Learning Approach for IoT
  Device Identification Based on Network Traffic Analysis}. In
  \bibinfo{booktitle}{\emph{Proceedings of the Symposium on Applied Computing}}
  \emph{(\bibinfo{series}{SAC'17})}. \bibinfo{publisher}{Association for
  Computing Machinery}, \bibinfo{address}{New York, NY, USA},
  \bibinfo{pages}{506–509}.
\newblock
\showISBNx{9781450344869}
\urldef\tempurl%
\url{https://doi.org/10.1145/3019612.3019878}
\showDOI{\tempurl}


\bibitem[\protect\citeauthoryear{Meidan, Sachidananda, Peng, Sagron, Elovici,
  and Shabtai}{Meidan et~al\mbox{.}}{2020}]%
        {meidan-1}
\bibfield{author}{\bibinfo{person}{Yair Meidan}, \bibinfo{person}{Vinay
  Sachidananda}, \bibinfo{person}{Hongyi Peng}, \bibinfo{person}{Racheli
  Sagron}, \bibinfo{person}{Yuval Elovici}, {and} \bibinfo{person}{Asaf
  Shabtai}.} \bibinfo{year}{2020}\natexlab{}.
\newblock \showarticletitle{A novel approach for detecting vulnerable IoT
  devices connected behind a home NAT}.
\newblock \bibinfo{journal}{\emph{Computers \& Security}}  \bibinfo{volume}{97}
  (\bibinfo{year}{2020}), \bibinfo{pages}{101968}.
\newblock
\showISSN{0167-4048}
\urldef\tempurl%
\url{https://doi.org/10.1016/j.cose.2020.101968}
\showDOI{\tempurl}


\bibitem[\protect\citeauthoryear{Msadek, Soua, and Engel}{Msadek
  et~al\mbox{.}}{2019}]%
        {iot-fingerprinting}
\bibfield{author}{\bibinfo{person}{Nizar Msadek}, \bibinfo{person}{Ridha Soua},
  {and} \bibinfo{person}{Thomas Engel}.} \bibinfo{year}{2019}\natexlab{}.
\newblock \showarticletitle{IoT Device Fingerprinting: Machine Learning based
  Encrypted Traffic Analysis}. In \bibinfo{booktitle}{\emph{2019 IEEE wireless
  communications and networking conference (WCNC)}}. \bibinfo{pages}{1--8}.
\newblock
\urldef\tempurl%
\url{https://doi.org/10.1109/WCNC.2019.8885429}
\showDOI{\tempurl}


\bibitem[\protect\citeauthoryear{Ngo, Nguyen, Le, and Nguyen}{Ngo
  et~al\mbox{.}}{2020}]%
        {mirai-2}
\bibfield{author}{\bibinfo{person}{Quoc-Dung Ngo}, \bibinfo{person}{Huy-Trung
  Nguyen}, \bibinfo{person}{Van-Hoang Le}, {and} \bibinfo{person}{Doan-Hieu
  Nguyen}.} \bibinfo{year}{2020}\natexlab{}.
\newblock \showarticletitle{A survey of IoT malware and detection methods based
  on static features}.
\newblock \bibinfo{journal}{\emph{ICT Express}} \bibinfo{volume}{6},
  \bibinfo{number}{4} (\bibinfo{year}{2020}), \bibinfo{pages}{280--286}.
\newblock
\showISSN{2405-9595}
\urldef\tempurl%
\url{https://doi.org/10.1016/j.icte.2020.04.005}
\showDOI{\tempurl}


\bibitem[\protect\citeauthoryear{Perdisci, Papastergiou, Alrawi, and
  Antonakakis}{Perdisci et~al\mbox{.}}{2020}]%
        {iot-finder}
\bibfield{author}{\bibinfo{person}{Roberto Perdisci}, \bibinfo{person}{Thomas
  Papastergiou}, \bibinfo{person}{Omar Alrawi}, {and} \bibinfo{person}{Manos
  Antonakakis}.} \bibinfo{year}{2020}\natexlab{}.
\newblock \showarticletitle{IoTFinder: Efficient Large-Scale Identification of
  IoT Devices via Passive DNS Traffic Analysis}. In
  \bibinfo{booktitle}{\emph{2020 IEEE European Symposium on Security and
  Privacy (EuroS\&P)}}. \bibinfo{pages}{474--489}.
\newblock
\urldef\tempurl%
\url{https://doi.org/10.1109/EuroSP48549.2020.00037}
\showDOI{\tempurl}


\bibitem[\protect\citeauthoryear{Pinheiro, Bezerra, and Campelo}{Pinheiro
  et~al\mbox{.}}{2018}]%
        {padding-2}
\bibfield{author}{\bibinfo{person}{Antônio~J. Pinheiro},
  \bibinfo{person}{Jeandro~M. Bezerra}, {and} \bibinfo{person}{Divanilson~R.
  Campelo}.} \bibinfo{year}{2018}\natexlab{}.
\newblock \showarticletitle{Packet Padding for Improving Privacy in Consumer
  IoT}. In \bibinfo{booktitle}{\emph{2018 IEEE Symposium on Computers and
  Communications (ISCC)}}. \bibinfo{pages}{00925--00929}.
\newblock
\urldef\tempurl%
\url{https://doi.org/10.1109/ISCC.2018.8538744}
\showDOI{\tempurl}


\bibitem[\protect\citeauthoryear{Pinheiro, {de M. Bezerra}, Burgardt, and
  Campelo}{Pinheiro et~al\mbox{.}}{2019}]%
        {pkt-sizes-identification}
\bibfield{author}{\bibinfo{person}{Antônio~J. Pinheiro},
  \bibinfo{person}{Jeandro {de M. Bezerra}}, \bibinfo{person}{Caio~A.P.
  Burgardt}, {and} \bibinfo{person}{Divanilson~R. Campelo}.}
  \bibinfo{year}{2019}\natexlab{}.
\newblock \showarticletitle{Identifying IoT devices and events based on packet
  length from encrypted traffic}.
\newblock \bibinfo{journal}{\emph{Computer Communications}}
  \bibinfo{volume}{144} (\bibinfo{year}{2019}), \bibinfo{pages}{8--17}.
\newblock
\showISSN{0140-3664}
\urldef\tempurl%
\url{https://doi.org/10.1016/j.comcom.2019.05.012}
\showDOI{\tempurl}


\bibitem[\protect\citeauthoryear{Pinheiro, Freitas~de Araujo-Filho,
  de~M.~Bezerra, and Campelo}{Pinheiro et~al\mbox{.}}{2021}]%
        {Levels-Padding}
\bibfield{author}{\bibinfo{person}{Antônio~J. Pinheiro},
  \bibinfo{person}{Paulo Freitas~de Araujo-Filho}, \bibinfo{person}{Jeandro de
  M.~Bezerra}, {and} \bibinfo{person}{Divanilson~R. Campelo}.}
  \bibinfo{year}{2021}\natexlab{}.
\newblock \showarticletitle{Adaptive Packet Padding Approach for Smart Home
  Networks: A Tradeoff Between Privacy and Performance}.
\newblock \bibinfo{journal}{\emph{IEEE Internet of Things Journal}}
  \bibinfo{volume}{8}, \bibinfo{number}{5} (\bibinfo{year}{2021}),
  \bibinfo{pages}{3930--3938}.
\newblock
\urldef\tempurl%
\url{https://doi.org/10.1109/JIOT.2020.3025988}
\showDOI{\tempurl}


\bibitem[\protect\citeauthoryear{Shahid, Blanc, Zhang, and Debar}{Shahid
  et~al\mbox{.}}{2018}]%
        {IoT-Devices-Recognition}
\bibfield{author}{\bibinfo{person}{Mustafizur~R. Shahid},
  \bibinfo{person}{Gregory Blanc}, \bibinfo{person}{Zonghua Zhang}, {and}
  \bibinfo{person}{Hervé Debar}.} \bibinfo{year}{2018}\natexlab{}.
\newblock \showarticletitle{IoT Devices Recognition Through Network Traffic
  Analysis}. In \bibinfo{booktitle}{\emph{2018 IEEE International Conference on
  Big Data (Big Data)}}. \bibinfo{pages}{5187--5192}.
\newblock
\urldef\tempurl%
\url{https://doi.org/10.1109/BigData.2018.8622243}
\showDOI{\tempurl}


\bibitem[\protect\citeauthoryear{Sivanathan, Sherratt, Habibi~Gharakheili,
  Radford, Wijenayake, Vishwanath, and Sivaraman}{Sivanathan
  et~al\mbox{.}}{2017}]%
        {Australian}
\bibfield{author}{\bibinfo{person}{Arunan Sivanathan}, \bibinfo{person}{Daniel
  Sherratt}, \bibinfo{person}{Hassan Habibi~Gharakheili}, \bibinfo{person}{Adam
  Radford}, \bibinfo{person}{Chamith Wijenayake}, \bibinfo{person}{Arun
  Vishwanath}, {and} \bibinfo{person}{Vijay Sivaraman}.}
  \bibinfo{year}{2017}\natexlab{}.
\newblock \showarticletitle{Characterizing and classifying IoT traffic in smart
  cities and campuses}. In \bibinfo{booktitle}{\emph{Proc. IEEE Conference on
  Computer Communications Workshops (INFOCOM WKSHPS)}}.
  \bibinfo{pages}{559--564}.
\newblock
\urldef\tempurl%
\url{https://doi.org/10.1109/INFCOMW.2017.8116438}
\showDOI{\tempurl}


\bibitem[\protect\citeauthoryear{Sridharan, Maiti, and Tippenhauer}{Sridharan
  et~al\mbox{.}}{2018}]%
        {anomaly-ml-2}
\bibfield{author}{\bibinfo{person}{Ragav Sridharan},
  \bibinfo{person}{Rajib~Ranjan Maiti}, {and} \bibinfo{person}{Nils~Ole
  Tippenhauer}.} \bibinfo{year}{2018}\natexlab{}.
\newblock \showarticletitle{WADAC: Privacy-Preserving Anomaly Detection and
  Attack Classification on Wireless Traffic}. In
  \bibinfo{booktitle}{\emph{Proceedings of the 11th ACM Conference on Security
  \& Privacy in Wireless and Mobile Networks}} (Stockholm, Sweden)
  \emph{(\bibinfo{series}{WiSec '18})}. \bibinfo{publisher}{Association for
  Computing Machinery}, \bibinfo{address}{New York, NY, USA},
  \bibinfo{pages}{51–62}.
\newblock
\showISBNx{9781450357319}
\urldef\tempurl%
\url{https://doi.org/10.1145/3212480.3212495}
\showDOI{\tempurl}


\bibitem[\protect\citeauthoryear{St{\"o}ber, Frank, Schmitt, and
  Martinovic}{St{\"o}ber et~al\mbox{.}}{2013}]%
        {smartphone-fingerprinting}
\bibfield{author}{\bibinfo{person}{Tim St{\"o}ber}, \bibinfo{person}{Mario
  Frank}, \bibinfo{person}{Jens Schmitt}, {and} \bibinfo{person}{Ivan
  Martinovic}.} \bibinfo{year}{2013}\natexlab{}.
\newblock \showarticletitle{Who do you sync you are?: smartphone fingerprinting
  via application behaviour}. In \bibinfo{booktitle}{\emph{Proc. 6th ACM
  conference on Security and privacy in wireless and mobile networks}}.
\newblock
\showISBNx{978-1-4503-1998-0}
\urldef\tempurl%
\url{https://doi.org/10.1145/2462096.2462099}
\showDOI{\tempurl}


\bibitem[\protect\citeauthoryear{Trimananda, Varmarken, Markopoulou, and
  Demsky}{Trimananda et~al\mbox{.}}{2020}]%
        {Ping-Pong}
\bibfield{author}{\bibinfo{person}{R. Trimananda}, \bibinfo{person}{Janus
  Varmarken}, \bibinfo{person}{A. Markopoulou}, {and} \bibinfo{person}{B.
  Demsky}.} \bibinfo{year}{2020}\natexlab{}.
\newblock \showarticletitle{Packet-Level Signatures for Smart Home Devices}. In
  \bibinfo{booktitle}{\emph{Network and Distributed Systems Security (NDSS)
  Symposium}}.
\newblock


\bibitem[\protect\citeauthoryear{Velan, Cermak, Celeda, and Drašar}{Velan
  et~al\mbox{.}}{2015}]%
        {survey}
\bibfield{author}{\bibinfo{person}{Petr Velan}, \bibinfo{person}{Milan Cermak},
  \bibinfo{person}{Pavel Celeda}, {and} \bibinfo{person}{Martin Drašar}.}
  \bibinfo{year}{2015}\natexlab{}.
\newblock \showarticletitle{A survey of methods for encrypted traffic
  classification and analysis}.
\newblock \bibinfo{journal}{\emph{International Journal of Network Management}}
   \bibinfo{volume}{25} (\bibinfo{date}{07} \bibinfo{year}{2015}).
\newblock
\urldef\tempurl%
\url{https://doi.org/10.1002/nem.1901}
\showDOI{\tempurl}


\bibitem[\protect\citeauthoryear{Wang, Kennedy, Li, Hudson, Atluri, Wei, Sun,
  and Wang}{Wang et~al\mbox{.}}{2020}]%
        {deep-1}
\bibfield{author}{\bibinfo{person}{Chenggang Wang}, \bibinfo{person}{Sean
  Kennedy}, \bibinfo{person}{Haipeng Li}, \bibinfo{person}{King Hudson},
  \bibinfo{person}{Gowtham Atluri}, \bibinfo{person}{Xuetao Wei},
  \bibinfo{person}{Wenhai Sun}, {and} \bibinfo{person}{Boyang Wang}.}
  \bibinfo{year}{2020}\natexlab{}.
\newblock \showarticletitle{Fingerprinting Encrypted Voice Traffic on Smart
  Speakers with Deep Learning} \emph{(\bibinfo{series}{WiSec '20})}.
  \bibinfo{publisher}{Association for Computing Machinery},
  \bibinfo{address}{New York, NY, USA}, \bibinfo{pages}{254–265}.
\newblock
\showISBNx{9781450380065}
\urldef\tempurl%
\url{https://doi.org/10.1145/3395351.3399357}
\showDOI{\tempurl}


\bibitem[\protect\citeauthoryear{Xiong, Sarwate, and Mandayam}{Xiong
  et~al\mbox{.}}{2018}]%
        {padding-3}
\bibfield{author}{\bibinfo{person}{Sijie Xiong}, \bibinfo{person}{Anand~D.
  Sarwate}, {and} \bibinfo{person}{Narayan~B. Mandayam}.}
  \bibinfo{year}{2018}\natexlab{}.
\newblock \showarticletitle{Defending Against Packet-Size Side-Channel Attacks
  in Iot Networks}. In \bibinfo{booktitle}{\emph{2018 IEEE International
  Conference on Acoustics, Speech and Signal Processing (ICASSP)}}.
  \bibinfo{pages}{2027--2031}.
\newblock
\urldef\tempurl%
\url{https://doi.org/10.1109/ICASSP.2018.8461330}
\showDOI{\tempurl}


\bibitem[\protect\citeauthoryear{Zhang, Meng, Liu, Zhang, Zhang, and Zhu}{Zhang
  et~al\mbox{.}}{2018}]%
        {anomaly-dfa}
\bibfield{author}{\bibinfo{person}{Wei Zhang}, \bibinfo{person}{Yan Meng},
  \bibinfo{person}{Yugeng Liu}, \bibinfo{person}{Xiaokuan Zhang},
  \bibinfo{person}{Yinqian Zhang}, {and} \bibinfo{person}{Haojin Zhu}.}
  \bibinfo{year}{2018}\natexlab{}.
\newblock \showarticletitle{HoMonit: Monitoring Smart Home Apps from Encrypted
  Traffic}. In \bibinfo{booktitle}{\emph{Proceedings of the 2018 ACM SIGSAC
  Conference on Computer and Communications Security}} (Toronto, Canada)
  \emph{(\bibinfo{series}{CCS '18})}. \bibinfo{publisher}{Association for
  Computing Machinery}, \bibinfo{address}{New York, NY, USA},
  \bibinfo{pages}{1074–1088}.
\newblock
\showISBNx{9781450356930}
\urldef\tempurl%
\url{https://doi.org/10.1145/3243734.3243820}
\showDOI{\tempurl}


\end{thebibliography}

\appendix

\section{The Active Adversary Model}\label{Anomaly Detection}
\subsection{Overview}

We now change our viewpoint to that of the defender trying to detect abnormal behavior of a device taken over by an active adversary. For that, we again use the packet-sizes distributions. 

Our goal is to detect whenever that happens. 
We used 3 different traces of 1 hour of traffic to develop and evaluate our methods - (1) normal-behavior-trace, (2) validation-trace and (3) test-trace.
In the learning phase we only use the normal-behavior-trace and the validation-trace. 

The normal-trace contains only normal-behavior traffic associated with the device. 
In order to check our anomaly detection methods, we use simulated Mirai traffic as captured from the real attack ~\cite{Attack-Vectors}, categorised into the three stages in the Mirai attacks discussed in Section \ref{The Data Corpora}.
For a given device trace and Mirai attack stage, 
we inject the attack packets into the regular traffic of the validation-trace of the device to simulate an active adversary exploiting the IoT network.
We use both the normal-trace and validation-trace to set a threshold to differentiate between windows with normal behavior traffic to windows containing Mirai abnormal packets.

In the test phase, we inject the Mirai traffic into the test-trace, and evaluate our models success-rates in detecting the anomalies. We repeat this experiment 10 times for each device.

We divided our validation-trace and test-trace into time windows and injected the malicious attack packets into 50\% of the windows in both traces.
We set the length of the time window to be 2 minutes long, so we have 30 windows in each trace. 

In this section we shall compare two detection methods: one based on the Local Outlier Factor (LOF) Algorithm, and the other one directly on the Jensen-Shannon distance (JSD)~\cite{kl-1,cosine/kl} between packet-sizes distributions: basic experimentation indicated that using the JSD as distance measure for both methods gives better anomaly-detection results than the cosine distance.
JSD is a metric that is bounded between 0 to 1, based on the Kullback-Leibler (KL) divergence between two normalized (i.e., with norm equal to 1) distributions' feature vectors which are defined on the same packet-sizes set $X$:
$$
KL(p||q)=\sum_{x \in X}p(x)\cdot \log{\frac{p(x)}{q(x)}}
$$

$$
JSD(u,v)=\sqrt{\frac{1}{2} KL(u|| \frac{u+v}{2}) + \frac{1}{2} KL(v|| \frac{u+v}{2})}
$$
KL divergence is asymmetric and is not defined if there exists a packet-size $x\in X$ such that $q(x)=0, p(x)\neq 0$. 
JSD distance which is a metric (in particular it's symmetric) fixes this and is always well-defined. 

\subsection{The Local Outlier Factor (LOF) Method}

The first anomaly detection method we used is the Local Outlier Factor (LOF) method, known for anomaly detection in many cases in the literature ~\cite{lof-2,lof-1}. 
We divide the normal-trace too into 30 time windows of 2 minutes long.
The feature vector of a time window in each trace is its packets-sizes distribution. 

To set the threshold we use the following process in the learning phase: we give a score to each validation-trace window, using the LOF algorithm as demonstrated in Figure \ref{lof-figure}.
Validation-trace windows that are further away from the normal behavior windows of the normal-trace will get higher scores.
We used the python library ~\cite{sklearn} \textit{sklearn.neighbors.LocalOutlierFactor } and specifically its attribute \textit{negative-outlier-factor }. Taking its opposite number gives the score.

After doing this, we have a score for every validation window. We then plot the ROC Curve for the validation trace, based on setting each time a different score as a threshold. 
Windows with score higher than the threshold are considered as abnormal, while windows with lower score are considered to be normal. We then take as our threshold the value that yields the best accuracy of classification. 

In our experimentation with this method of setting the threshold, we got a perfect ROC Curve with Area Under Curve  (AUC) of 1 on the validation trace, and the chosen threshold also incurs Equal Error Rate (EER) of 0,
which means that our method can differentiate perfectly between the two classes of windows in the validation trace.

Last, we use the chosen threshold: each test-trace window gets a score in a similar way and is classified as abnormal if its score is higher than the chosen threshold, and normal otherwise. We got Precision of 100\% over all the devices, i.e., we didn't have any false alarms.
Table \ref{Anomaly Detection - Results} shows that we have a Recall of at least 93\% for each device.

\subsection{The Jensen-Shannon (JS) Method}

\begin{figure}
     \centering
     \begin{subfigure}[t]{0.45\textwidth}
         \centering
         \captionsetup{justification=centering}
         \includegraphics[width=\textwidth]{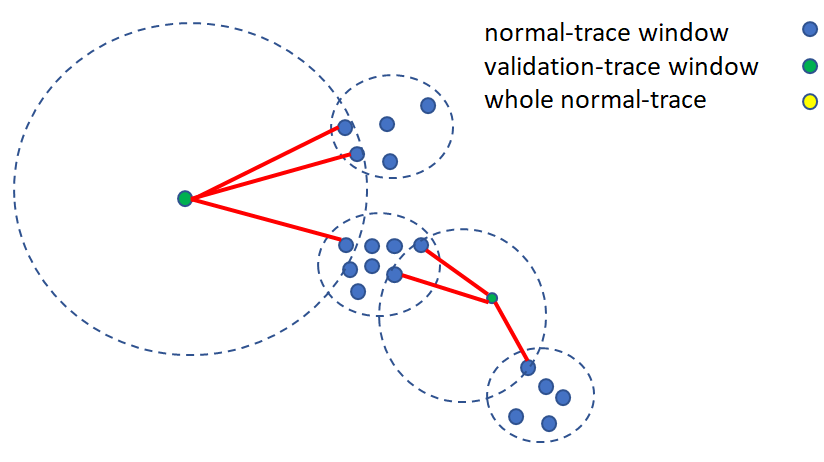}
         \caption{LOF Method}
         \Description{Figure showing how lof learning process work}
         \label{lof-figure}
     \end{subfigure}
     \hfill
     \begin{subfigure}[t]{0.45\textwidth}
         \centering
         \captionsetup{justification=centering}
         \includegraphics[width=\textwidth]{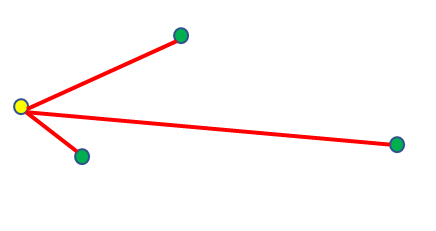}
         \caption{JS Method}
         \Description{Figure showing how jensen-shannon learning process work}
         \label{js-figure}
     \end{subfigure}
        \caption{Scoring validation-trace windows}
        \Description{Showing how each method scores its validaion windows}
        \label{Scoring validation-trace windows}
\end{figure}

In this method we use directly our Jensen-Shannon distance metric to set the thresholds.
This time we considered the whole learning day as a whole packet-sizes distribution rather than dividing it to time windows like in the LOF method. 
Similarly to the LOF method, we give a score to each validation-trace window. Figure \ref{js-figure} shows that each window gets as score the value of the jensen-shannon distance between it and the whole normal-trace.
Validation-trace windows with higher distance from the whole normal-trace get higher score.
Like before, we plot the ROC curve and choose the best threshold.
We got again a ROC Curve with AUC of 1 and EER of 0, which means that this method can also distinguish perfectly between the two types of windows.

As before, we use the chosen threshold in order to classify each window from the test-trace.
We got again Precision of 100\% over all the devices, so we didn't have any false alarms.
Table \ref{Anomaly Detection - Results} shows that we have a Recall of at least 89\% for each device.

\subsection{Comparing The Methods}

Both methods did not have any false alarms caused by the detection method (i.e., never alerted about anomaly window when it wasn't the case) and had Precision of 100\% for all the devices.
As displayed in Table ~\ref{Anomaly Detection - Results}, the LOF method has slightly better rates of Recall than the JS method, and both have at least 89\% Recall over all the devices.   
Abnormal windows' distributions are significantly different than normal windows' distributions because the attack packets injected have common sizes of Mirai attacks that repeat themselves in a high rate (e.g. DNS attack). Thus, our methods managed to differentiate between the distributions associated with the normal and abnormal windows.

\begin{table}[t]
\centering
\caption{Anomaly Detection - Results}
\begin{tabular}{|M{2.6cm}|M{2.6cm}|M{2.6cm}|}
\hline

Device Name &Recall - JS Method
(\%)&Recall -  LOF Method
(\%)\\\hline
TP-Link Plug & 100 & 100  \\\hline
Belkin Motion & 95 & 100  \\\hline
Amazon Echo & 94 & 96  \\\hline
Netatmo Weather & 89 & 94  \\\hline
Samsung Camera & 100 & 100  \\\hline
HP Printer & 89 & 93  \\\hline
TP-Link Camera & 96 & 100  \\\hline
Amazon Plug & 89 & 95  \\\hline
D-Link & 90 & 100  \\\hline
Rachio Sprinkler & 89 & 94  \\\hline
Ring Alarm & 93 & 94  \\\hline
Roomba & 93 & 93  \\\hline
TP-Link Bulb & 94 & 94  \\\hline
Wemo Insight Plug & 93 & 100  \\\hline

\end{tabular}
\label{Anomaly Detection - Results}
\end{table}

\end{document}